\documentclass[%
 reprint,
 amsmath,amssymb,
 aps,nofootinbib,   
]{revtex4-1}
\usepackage{bm}
\PassOptionsToPackage{linktocpage}{hyperref}
\usepackage[hyperindex,breaklinks]{hyperref}
\usepackage{enumitem}
\usepackage{slashed}

\renewcommand{\theta}{\vartheta}

\renewcommand{\vec}[1]{\ensuremath{\boldsymbol{#1}}}

\newcommand{\bra}[1]{\ensuremath{\left< #1\,\right|}}
\newcommand{\ket}[1]{\ensuremath{\left|\, #1\right>}}

\usepackage{array}
\usepackage{mathtools}

\usepackage{etoolbox}

\begin{document} 

\title{Entropy Bound and Unitarity of Scattering Amplitudes}

\author{Gia Dvali} 
\affiliation{%
Arnold Sommerfeld Center, Ludwig-Maximilians-Universit\"at, Theresienstra{\ss}e 37, 80333 M\"unchen, Germany, 
}%
 \affiliation{%
Max-Planck-Institut f\"ur Physik, F\"ohringer Ring 6, 80805 M\"unchen, Germany
}%

\date{\today}

\begin{abstract}

  We establish that unitarity of scattering amplitudes imposes 
 universal entropy bounds. 
 The maximal entropy of a self-sustained quantum field
object of radius $R$
 is equal to its surface area and at the same time to the inverse running coupling $\alpha$ evaluated at the scale $R$. The saturation of these entropy bounds is in one-to-one  correspondence with the non-perturbative saturation of unitarity by 
$2 \rightarrow N$ particle scattering amplitudes at the point of optimal truncation.  
These bounds are more stringent
than Bekenstein's bound and in a consistent theory 
all three get saturated simultaneously.
This is true for all known entropy-saturating objects 
such as solitons, instantons, baryons, oscillons, black holes or simply lumps of classical fields.  We refer to these collectively as {\it saturons}
and show that in renormalizable theories they behave in all other respects like black holes. 
Finally, it is argued that the confinement in $SU(N)$ gauge theory 
can be understood as a direct consequence of the entropy bounds
and unitarity.

\end{abstract}

\maketitle
\section{Introduction}

The purpose of this paper is to 
show that unitarity of scattering amplitudes imposes the following  {\it universal} non-perturbative upper bounds on the entropy  
of the system. 

 \begin{itemize}
  \item {\bf The area-law entropy bound: \\
  
   The maximal entropy of any 
  self-sustained quantum field theoretic object localized within
  a sphere of radius $R$ is equal to the area of the sphere measured 
  in units of the relevant Goldstone decay constant $f$: 
   \begin{equation} \label{GiaA}
       S_{\rm max} \,  =  \frac{\rm Area}{f^{-2}}  \,.  
       \end{equation}   

  \item  The inverse-coupling entropy bound: \\
  
   The maximal entropy of any 
  self-sustained quantum field theoretic object localized within
  a sphere of radius $R$ is equal to the inverse of the running coupling
 $\alpha(q)$ of the 
 relevant long-range interaction evaluated at the scale of 
 momentum-transfer $q = \frac{1}{R}$.
 
  \begin{equation} \label{GiaC}
       S_{\rm max} \,  = \, {1 \over \alpha} \,.   
       \end{equation} }  
 
\end{itemize} 
  
   We shall argue that a violation of the above bounds leads to 
   a {\it non-perturbative} violation of unitarity.  \\

  The  
foundation for this connection was already laid down 
in previous articles \cite{Gia1,Gia2}.
 Namely, it was observed there that entropy of a self-sustained 
 field theoretic object such as soliton or a baryon of 
 mass $M$ and radius $R$ saturates (\ref{GiaA}) and (\ref{GiaC})
 simultaneously with Bekenstein's entropy bound 
 \cite{BekBound}, 
\begin{equation} \label{Bek1}
    S_{\rm max} =  2\pi MR \, .  
\end{equation}
This happens {\it exclusively}  when 
the theory saturates unitarity.
That is, the following relations emerge. \\
 
First, the maximal entropy  is always  
equal 
to the surface area of the object, 
  measured  in units  of the decay constant $f$ of the Goldstone field,
 as given by  (\ref{GiaA}). 
  This Goldstone mode is {\it universally}  present due to the fact that 
  any localized field configuration breaks spontaneously set of symmetries, which obviously include Poincare translations.  
 However, there also emerge the  Goldstone mode(s) corresponding to the breaking of internal symmetries. 
 This shall become clear below. 
  \\
  
Secondly, the same maximal entropy is equal to an inverse of the running coupling $\alpha$ evaluated at the scale 
$q=1/R$,  as described by (\ref{GiaC}).  Of course, what matters 
is the interaction with the range that covers $R$.
Note, when the scale $R$ separates two different regimes, the equation 
(\ref{GiaC}) must be satisfied from both sides. For example,  
in case of a baryon of size $R$, it is satisfied both by gluons and by pions. 
\\ 

    Thus, in \cite{Gia1,Gia2}  the entropy bound attained by various objects was observed to satisfy the following relation,   
   \begin{equation} \label{Gia}
       S_{\rm max} \,  = MR = {1 \over \alpha} =  \frac{\rm Area}{f^{-2}} \,.  
     \end{equation}
 (Throughout the paper, the order-one numeric factors shall be explicitly shown only when they are important.) \\

  From here, the following natural questions emerge:
  \begin{itemize}
  \item Are the three 
  bounds  (\ref{GiaA}), (\ref{GiaC}) and (\ref{Bek1}) equivalent? 
  \item  And if not, which of them is more fundamental?
\end{itemize}

  The main goal of the present paper is to understand the independent fundamental meanings of the area-law (\ref{GiaA}) and the inverse-coupling (\ref{GiaC}) 
 entropy bounds and their connection to unitarity. 
 First, we shall achieve this by analysing scattering amplitudes. 
 Secondly, we shall construct explicit renormalizable theories in which 
 the saturation of the three different bounds can be monitored 
 in various parameter regimes.  \\

 The first part of our message is to establish an universal connection 
 between the bounds (\ref{GiaA}) and (\ref{GiaC}) and scattering amplitudes. 
Namely, there exists a one-to-one correspondence 
between the saturation of (\ref{GiaA}) and (\ref{GiaC})
 by an arbitrary field theoretic entity 
- irrespectively whether of Lorentzian or Euclidean signature - 
and non-perturbative saturation of unitarity by a set 
of $2\rightarrow n$ amplitudes with $n=\frac{1}{\alpha}$ 
at momentum-transfer $q=\frac{1}{R}$. 
This saturation is non-perturbative and cannot be removed by resummation.  \\

Surprisingly, the bounds (\ref{GiaA}) and (\ref{GiaC}) turn out to be 
 more stringent than the Bekenstein
bound (\ref{Bek1}). 
As we shall see, in some situations these bounds can be violated even when the 
Bekenstein bound (\ref{Bek1}) is still respected. 
Such examples are immediately killed by unitarity. 
This is because the bounds (\ref{GiaA}) and (\ref{GiaC}) control
the saturation of unitarity by the scattering amplitudes.
On the other hand, in all  examples known to us, 
the saturation of the bounds (\ref{GiaA}) and (\ref{GiaC}) 
automatically leads to the saturation of the bound 
(\ref{Bek1}).  Therefore, the saturation of the two former bounds appears 
to provide the necessary and sufficient condition for 
reaching the maximal entropy permitted by the consistency of the theory.
Thus, in a consistent theory at the saturation point the entropy 
satisfies the triple equation (\ref{Gia}).   
 \\

A natural physical interpretation of the above 
amplitudes at the saturation point 
is that they describe a creation of $n$-particle  composite 
object. This object saturates the entropy bounds
(\ref{GiaA}) and (\ref{GiaC}) and correspondingly satisfies (\ref{Gia}). 
We shall refer to such objects as {\it saturons}. The process thus 
schematically can be presented as a creation of a classical object
in a two-particle scattering, 
\begin{equation} \label{Asaturon}
  2 \rightarrow n = \,  {\rm  saturon} \, .
 \end{equation}   
 The reason why the cross-section of such a process is not exponentially 
suppressed is that the saturon exhausts all possible final states in
the given kinematic regime.  So in this sense saturons effectively 
provide the mechanism of  {\it classicalization}  of the scattering amplitude \cite{Cl1}.  
Of course, explaining how this happens is one of the central points of our paper. \\

 However, the above should not create a false impression that it is easy 
 to produce a saturon in a high energy scattering experiment.  
 Although, at its mass-threshold the saturon's cross section saturates unitarity  
 at the expense of its
 maximal entropy, there is a price to pay. 
 It comes in form of a very narrow
($\frac{\Delta E}{E} \sim \alpha$)  ``window of opportunity" for the choice of the center of mass energy $E$ of the initial state. 
Due to this, in order 
for saturons to play a role in UV-completion of the theory,
they must fill an almost continuous mass spectrum. 
This is possible if the theory possess a non-trivial fixed point. 
In such a case, saturons can play an interesting role both 
in UV-completion as well as in collider phenomenology.  
\\

From the point of view of fundamental physics, one of the
implications of the bounds (\ref{GiaA}) and 
(\ref{GiaC}) is  to put  phenomena such as confinement 
in a new light. Namely,   
it was already suggested in \cite{Gia1} that 
confinement in $SU(N)$ gauge theory  can be viewed as 
a built-in defence mechanism against violations of the entropy bounds.
 Here, we provide more evidence for this.   Namely, 
we consider an example 
presented in \cite{Gia2} of $SU(N)$ gauge theory 
in which the entropy bounds (\ref{GiaA}) and 
(\ref{GiaC}) are saturated by an instanton.  
We show that this saturation is mapped on the saturation of unitarity by a set $2 \rightarrow N$-gluon amplitudes. 
From here it  is evident that in order not to violate these bounds 
the theory must become confining at large distances.  That is, 
without confinement there is no visible mechanism that would prevent 
such a violation at some IR scale.  \\

Analogously, when quarks are included,  the theory resists against violation of the bounds (\ref{GiaA}) and (\ref{GiaC}) by 
baryons. Namely, a baryon saturates  both entropy bounds 
when the number of the quark flavors is of the same order as the number of colors.  The baryon entropy in this limit is given by its area measured 
in units of the pion decay constant \cite{Gia1}.  Simultaneously, the 
$2 \rightarrow N$ pion cross section saturates unitarity.  
 In this case, the  violation of the bounds (\ref{GiaA}) and (\ref{GiaC}) would render the theory asymptotically {\it not free} and thus inconsistent in UV. \\

Finally, an important message of the present paper is the  
understanding of black holes and the saturons of renormalizable 
theories as the representatives of the same saturon family. In order to make the parallels maximally sharp, 
we construct an explicit renormalizable theory which contains 
saturons. These are the solitonic vacuum bubbles. In the interior of the 
 bubble $N$ distinct gapless Goldstone modes are localized.   
   These gapless modes endow the bubble 
 with a large micro-state entropy. We then show that at the point when the bubble saturates the entropy bounds (\ref{GiaA}) and (\ref{GiaC}), 
 the corresponding amplitudes saturate unitarity. 
 So, the bubble becomes a saturon. 
 At this point, all its properties become
 identical to the known properties of a black hole. \\
 
 For example, both the renormalizable 
 saturon and a black hole obey the relation 
 (\ref{Gia}).  Here, we must remember that for a black hole 
 $f = M_P$, where $M_P$ is the Planck mass. 
Indeed, first, $M_P$ represents the graviton decay constant. 
Secondly, the Goldstone boson of a translation symmetry
that is spontaneously broken
 by a black hole, is the graviton itself. 
This immediately shows that  the famous
 Bekenstein-Hawking entropy \cite{BekE} satisfies 
 the relation (\ref{Gia}).   
  Next, just like a black hole, in the semi-classical limit 
 ($N = \infty$) the non-gravitational saturon possesses an {\it information horizon}.   
 It emits particles in a way that is strikingly similar to Hawking's emission. 
 In particular, the information stored in the saturon's interior cannot be
 decoded by analysing the emitted radiation. 
 In contrast, for finite $N$, the saturon bubble does release  
 information albeit  very slowly. The time-scales are identical to the ones 
 that are commonly attributed to a black hole.   
 Finally, both a black hole and a non-gravitational saturon saturate unitarity in respective 
 multi-particle scatterings. 
 This features are universal and independent on a particular nature of 
 a saturon. So they are  shared by saturons 
in other renormalizable theories.  \\

 The natural interpretation of the above striking connection is that 
a black hole of size $R$ represents a saturated state of the soft gravitons of wavelength $R$,   
as this has been long advocated by the black hole $N$-portrait \cite{NP}. 
In this paper the relation (\ref{GiaC}) for black holes has already been noticed. 
This relation was used there as a guiding principle for establishing 
the similarity between black holes and other saturated states such as 
Bose-Einstein condensates at criticality.  The present paper 
reinforces this view.

  \section{Entropy of a Lump} 
  
 Before moving to amplitudes, we shall establish mapping 
 between localized field  theory configurations with Lorentzian 
 signature,  such as solitons or lumps,  and $n$-particle states.
  We explain why for such objects the bound (\ref{Gia}) 
 holds.  \\ 
   
  \subsection{Lump as multi-particle state} 
  
Consider degrees of freedom described by creation/annihilation operators $\hat{a}_j(\vec{k})^\dagger, 
\hat{a}_j(\vec{k})$.  Here the label $\vec{k}$ refers to momentum, 
whereas $j=1,...,N$ is the species label describing different 
spin and internal states.  
 For example, $j$ can denote sets of color 
or flavor indexes.  We shall assume that operators obey the standard 
bosonic commutation relations, 
   $ [\hat{a}_i(\vec{k}),\hat{a}_j(\vec{k'})^{\dagger}] = \delta_{ij}\delta_{
    \vec{k}\vec{k'}} \,,
  [\hat{a}_i(\vec{k}),\hat{a}_j(\vec{k'})]  = 0 $. 
  That is, $\hat{a}_j(\vec{k})$ represent different physical modes of a bosonic quantum field $\hat{\phi}_j$, 
   \begin{equation} \label{expansion}
 \hat{\phi}_j  = \sum_{\vec{k}} {1\over \sqrt{\omega_{\vec{k}}}}
 \left ( {\rm e}^{i\vec{k}\vec{x}} \hat{a}_j(\vec k)  +  {\rm e}^{-i\vec{k}\vec{x}} \hat{a}_j(\vec k)^{\dagger} \right ) \,.  
 \end{equation} 
   This field can either be fundamental or represent an effective description of some 
  more fundamental theory.  For example, $\hat{\phi}_j$ may represent 
  the low energy fluctuations of quark-anti-quark condensate in QCD.  
  We shall also assume that the effective Hamiltonian is invariant 
  under an internal symmetry  ${\mathcal  G}$
 that acts on the label $j$.  Again, this symmetry can be either emergent 
 or be fundamental. \\

  Next, we shall denote by $\alpha$ the strength of an effective four-boson interaction,
  \begin{equation}\label{4point}
      \alpha \,  (\hat{\phi}_i\hat{\phi}_i)( \hat{\phi}_j\hat{\phi}_j)   \, + \, ... \,,
  \end{equation} 
   The above notation is highly schematic.    
 Throughout the paper we shall assume the coupling $\alpha$  to be 
 weak. In fact, defining the analog of the 't Hooft coupling,
 \begin{equation} \label{thooft} 
 \lambda_t \equiv  \alpha N \, ,  
   \end{equation} 
 our methods shall be most reliable in  the limit, 
 \begin{equation} \label{tlimit}
 \alpha \rightarrow 0,~\lambda_t={\rm finite} \,.   
\end{equation}
This is analogous to 't Hooft's limit \cite{planar}.
 \\

 Now, we wish to focus on states in which modes of certain momentum $\vec{k}$  
 are highly occupied 
 \begin{equation}\label{nstate}
 \ket{n}_{\rm micro} = 
 \prod_{j=1}^N {(\hat{a}_j(\vec{k})^{\dagger})^{n_j} \over \sqrt{n_j!}} \ket{0}\, , 
\end{equation}
where $n$ refers to a  total occupation number, 
\begin{equation}\label{totaln}
n = \sum_{j=1}^{N} n_j   \,.
\end{equation}
This number will be assumed to be very large.  We shall refer 
to such states as {\it micro-states}. This is because they are distinguished solely by different microscopic distributions of the total occupation number $n$ among the
$j$-species.  And, in the limit (\ref{tlimit}) they become indistinguishable.
 Such states therefore describe different micro-states of the same macro-state $\ket{n}$. \\

Obviously, in such a state the wave-functions of $n$ bosonic modes overlap, similarly to what happens in Bose-Einstein condensates. 
  It is therefore useful to introduce a concept of the 
{\it collective coupling} defined as, 
  \begin{equation} \label{collective} 
 \lambda_c \equiv  \alpha n \, .   
   \end{equation} 
 Again, our analysis is most reliable in the following   
 {\it double-scaling}  limit, 
 \begin{equation} \label{climit}
 \alpha \rightarrow 0,~\lambda_c={\rm finite} \,.   
\end{equation} 
Despite the superficial similarity between $\lambda_c$ and $\lambda_t$,
the two couplings are 
physically very different.  It is enough to note that the 't Hooft coupling 
$\lambda_t$ is a parameter of the 
{\it theory}, whereas the collective coupling $\lambda_c$ is a parameter of the {\it state}.  
   Despite this difference, as we shall see, the two couplings become comparable and critical on the states that saturate the entropy bounds
 (\ref{GiaA}), (\ref{GiaC}) and  (\ref{Gia}). \\
   
   Now, using the number-eigenstates (\ref{nstate}), we can form the coherent states  that represent classical field-configurations localized within certain characteristic radius 
  $R$.  They have a form,
  \begin{equation}\label{sol}
 \ket{sol} = {\rm e}^{\sum_{\vec{k}}\sum_{j=1}^N \sqrt{n_j(\vec{k})} (\hat{a}_j(\vec{k})^{\dagger}
 - \hat{a}(\vec{k})_j)} \ket{0} \,,
\end{equation}
with 
\begin{equation}\label{totalnk}
\sum_{j=1}^{N} \sum_{\vec{k}} n_j(\vec{k}) \, =\,  n \gg 1 \,,
\end{equation}
 where $n_j(\vec{k})$-s are sharply peaked around the characteristic momentum $|\vec{k}| \sim {1 \over R} \equiv q$. Obviously, the corresponding 
 classical field is described by the expectation value,
 \begin{equation} \label{expansion}
 \phi_j =  \bra{sol} \hat{\phi}_j \ket{sol}\,,
\end{equation}
of the quantum field.  We shall refer to such a state as a {\it lump} or a 
{\it soliton}.  
Of course, such a field configuration in general
depends on time.  It evolves both classically  
as well as quantum mechanically. Since the quantum coupling $\alpha$ is weak, the classical (mean field) evolution is valid for sufficiently long time. 
We are interested in field configurations that spread-out from the initial localization on time-scales $t \gg R$. 
This constraint does not apply to internal oscillations of the lump, as long as they  stay localized within 
the radius $R$. 
At weak coupling, this requirement is satisfied by most of the self-sustained solitonic configurations.  
The  condition for self-sustainability will be derived below.  \\

Under such conditions, the localized classical field configuration, 
$\phi_{sol}$,  can be treated as $n$-particle state of 
characteristic momenta $\sim q = 1/R$, each contributing 
$\sim q$ into the energy of the lump. The total 
energy therefore is,
\begin{equation} \label{energyn}
  E \sim {n \over R} \,.   
\end{equation}

Now, assuming that at distances $\sim R$ the 
interaction is attractive, let us estimate the number of constituents required 
for
creating a self-sustained bound-state.  
This can be done by balancing the kinetick energy of each quantum, 
$E_{kin} \sim \frac{1}{R}$, against the attractive potential 
energy from the rest. The latter goes as $E_{pot} \sim \frac{\alpha n}{R}$. 
This gives the equilibrium condition, 
\begin{equation} \label{balance} 
{\rm {\it Critical~balance}:}~~  \lambda_c =  \alpha n \sim  1 \,.  
\end{equation} 
We thus learn that the self-sustained configuration is reached when the collective coupling
$\lambda_c$ is order one, or equivalently, when $n \sim \frac{1}{\alpha}$.   
Inserting this relation in (\ref{energyn}), we get for the energy of the bound-state,  
 \begin{equation} \label{energySol}
  E_{\rm sol}  \sim {q \over \alpha} \sim  {1 \over \alpha R} \,.   
\end{equation} 
  The latter is a well-known relation between the 
  energy of a soliton and its size. \\
  
  Note,  of corse, in general, in a self-sustained bound-state, the particles do not strictly satisfy the dispersion relation 
$\omega_{\vec{k}} = \sqrt{m^2 + |\vec{k}|^2}$ with $m$ being a mass of a free particle.  That is, the operators $\hat{a}_j(\vec k)$ of the bound-state are related 
with analogous operators of free asymptotic quanta   
by a non-trivial Bogoliubov transformation. 
However, in the 
regime  (\ref{balance}) at large-$n$ this difference is unimportant for our purposes. 
In this regime, the self-sustained states can consistently be mapped 
on the scattering amplitudes.  \\

 \subsection{Inverse-coupling $=$ area-Law $=$ unitarity}     

We now wish to derive the entropy of the lump and establish 
for which values of parameters it saturates the bounds (\ref{GiaA}) and 
 (\ref{GiaC}).  
For this, we need to count the number of degenerate 
micro-states.  As already noted, the states (\ref{nstate}) (or  
 (\ref{sol})) represent particular micro-states belonging to one and the same classical macro state. This is due to the following reasons. 
First, such states form large representations under  
the symmetry ${\mathcal  G}$ that acts on the label $j$. 
Secondly, because the quantum coupling $\alpha$ is vanishingly small, 
the time-scale for differentiating between individual 
``colors" or ``flavors" is macroscopically large. Correspondingly, such states are {\it classically indistinguishable}. \\

Thus, the number of degenerate micro-states is given by the dimensionality of 
representation that they form under the symmetry group
${\mathcal  G}$. This dimensionality is easy to estimate. 
For example, in the simplest case of a symmetric wave-function, 
$n_j$-s can assume arbitrary values 
subject to 
 the constraint (\ref{totaln}) (or (\ref{totalnk})).   Therefore, the number of micro-states is given by the following binomial coefficient: 
  \begin{equation}\label{Nstates} 
 n_{\rm st} \simeq  \begin{pmatrix}
    n + N \\
     N  
\end{pmatrix} \, = c_N  \,  
\left ((1 + \frac{\lambda_t}{\lambda_c})^{\lambda_c}
 (1 + \frac{\lambda_c}{\lambda_t})^
 {\lambda_t} \right )^{\frac{1}{\alpha}} \, , 
  \end{equation}
  where we have used the Stirling approximation for  large $N=\frac{\lambda_t}{\alpha}$ and 
 $n=\frac{\lambda_c}{\alpha}$. 
Notice, the coefficient $c_N \simeq \sqrt{\frac{1}{2\pi}(N^{-1} +n^{-1})}$ can be replaced by one without any loss of information. 
This is the benefit of working at large $N$ and at the saturation point.  
 Since we shall take advantage of this fact throughout the paper, 
 we shall explain it briefly here. \\
 
 The trick is that the 
 saturation values of $\lambda_t$ and $\lambda_c$ 
 are determined by matching the quantities that are exponentially 
 sensitive to $N$ and $n$  (equivalently, to $\alpha^{-1}$).  
  Therefore, the coefficients such as $c_N$, that exhibit  power-law dependence 
  on $N$ and $n$, play essentially no role in it. Such quantities 
  correct the saturation value of $\lambda_t$ only by 
  the amount $\sim \frac{\ln(N)}{N}$ which vanishes in the 't Hooft limit (\ref{tlimit}). 
  Therefore, all such coefficients can be set equal to one without 
  compromising our analysis. \\

 Then, taking the collective coupling at the critical value 
 $\lambda_c=1$,  the number of states becomes
   \begin{equation}\label{NstatesC} 
 n_{\rm st}  \simeq  \,  
\left ((1 + \lambda_t)
 (1 + \frac{1}{\lambda_t})^
 {\lambda_t} \right )^{\frac{1}{\alpha}} \, . 
  \end{equation} 
The corresponding entropy of the soliton/lump is, 
\begin{equation} \label{EntropyS} 
S  = \ln(n_{\rm st}) \simeq {1 \over \alpha} \ln\left ((1 + \lambda_t)
 (1 + \frac{1}{\lambda_t})^
 {\lambda_t} \right )
\, .  
\end{equation} 
This entropy saturates the bound (\ref{GiaC}) for,
\begin{equation} \label{Saturation} 
{\rm {\it Entropy~saturation}:}~~ \lambda_t \simeq 0.54\,.  
\end{equation} 
Of course, what matters is that the critical  't Hooft coupling is order one. 
However, the above numerical value obtained for  
$\lambda_c =1$ will be useful as a reference point
for the later estimates.  As a consistency check, notice that the 
actual value of $c_N$ corrects (\ref{Saturation}) 
by the amount $\sim \frac{\ln(N)}{N}$ and is negligible. 
\\
 
Thus, we discover that the $n$-particle state, 
describing a self-sustained classical soliton/lump, saturates 
the entropy bound  (\ref{GiaC}) when the 't Hooft and collective couplings are both of order one, 
  \begin{equation} \label{lambdas} 
\lambda_c  \sim  \lambda_t  \sim 1 \, . 
\end{equation} 
As already pointed out in \cite{Gia1,Gia2}, through the above equation, the saturation of entropy is correlated with the saturation 
 of unitarity.  The depth of this correlation will be explored throughout the 
 paper. 
\\

  Now, following \cite{Gia1,Gia2}, it is easy to see that at the saturation point the entropy becomes
  equal to an area of the soliton/lump in units of the Goldstone decay constant
  $f$.  Let us therefore determine the latter. 
    The localized classical field configuration $\phi$ breaks spontaneously 
    both the space-translations as well as the internal 
     symmetries. The order parameter of breaking the translation invariance 
is $\nabla \phi \sim \frac{1}{R^2 \sqrt{\alpha}}$. 
    Consequently, the decay constant of the corresponding Goldstone 
    fields is
    \begin{equation}\label{Gold}
  f = \frac{1}{R \sqrt{\alpha}} = \frac{\sqrt{N}}{R} \,.
 \end{equation} 
 Notice, the above 
 expression also determines the decay constants of the Goldstone modes 
 of spontaneously broken internal symmetries. These are the symmetries 
  under which the lump/soliton transforms non-trivially. 
 Previously, they were schematically denoted by ${\mathcal  G}$. 
   The explicit examples will be constructed below. \\

      It is now obvious that the entropy (\ref{EntropyS}) at the 
   saturation point of the bound (\ref{GiaC})    
  can be written as, 
    \begin{equation} \label{AAA} 
S_{\rm max}  = {1 \over \alpha}  = (Rf)^2 = \frac{\rm Area}{f^{-2}}\,. 
\end{equation} 
Thus, the areal-law bound (\ref{GiaA}) is saturated
simultaneously with  (\ref{GiaC}).
As already stressed in \cite{Gia1,Gia2}, this is strikingly similar 
to a black hole entropy with the role of the Planck mass 
played by $f$.   \\

Thus, we discover that the saturation of the inverse-coupling entropy 
bound (\ref{GiaC}) takes place together with the saturation of the 
area-law bound (\ref{GiaA}). It is very important that this happens when the value of  the 't Hooft coupling is order one
(\ref{lambdas}).   
This fact is the key for connecting the saturation of the above entropy  bounds to unitarity. \\

Notice, the saturation of the bounds (\ref{GiaA}) and (\ref{GiaC}) 
implies the saturation of the Bekenstein bound (\ref{Bek1}).   
This can be  seen easily by inserting (\ref{energySol}) in the  
Bekenstein formula (\ref{Bek1}).  We get 
\begin{equation} \label{Gia1} 
S_{\rm Bek} = ER = {1 \over \alpha} \, . 
\end{equation} 
Thus, a self-sustained quantum field theoretic system 
with a single characteristic localization scale $R$ satisfies (\ref{Gia}). 
This is exactly the result obtained in \cite{Gia1,Gia2}. \\

However, the converse is not true in general. That is, a  satisfaction of the 
Bekenstein bound (\ref{Bek1}) does not guarantee the satisfaction of the 
bounds (\ref{GiaA}) or (\ref{GiaC}). However, such examples 
violate unitarity and, therefore, are inconsistent.  Thus, the areal-law 
and the inverse-coupling bounds turn out to be more restrictive
than the Bekenstein bound. We conclude that in a consistent theory all three bounds must be saturated together (\ref{Gia}).  In all  
examples known to us this proves to be 
the case.

 \section{Connection with Amplitudes}

 The equation (\ref{EntropyS}) tells us that the classical lump 
 saturates the entropy bounds (\ref{GiaA}) and (\ref{GiaC}) 
and satisfies  (\ref{Gia}) when the 't Hooft coupling
 $\lambda_t$  equals to the critical value (\ref{lambdas})
 (or more explicitly, (\ref{Saturation})). 
  We now wish to connect this phenomenon to the saturation of unitarity by certain scattering 
  amplitudes. \\
  
\begin{figure}
 	\begin{center}
        \includegraphics[width=0.53\textwidth]{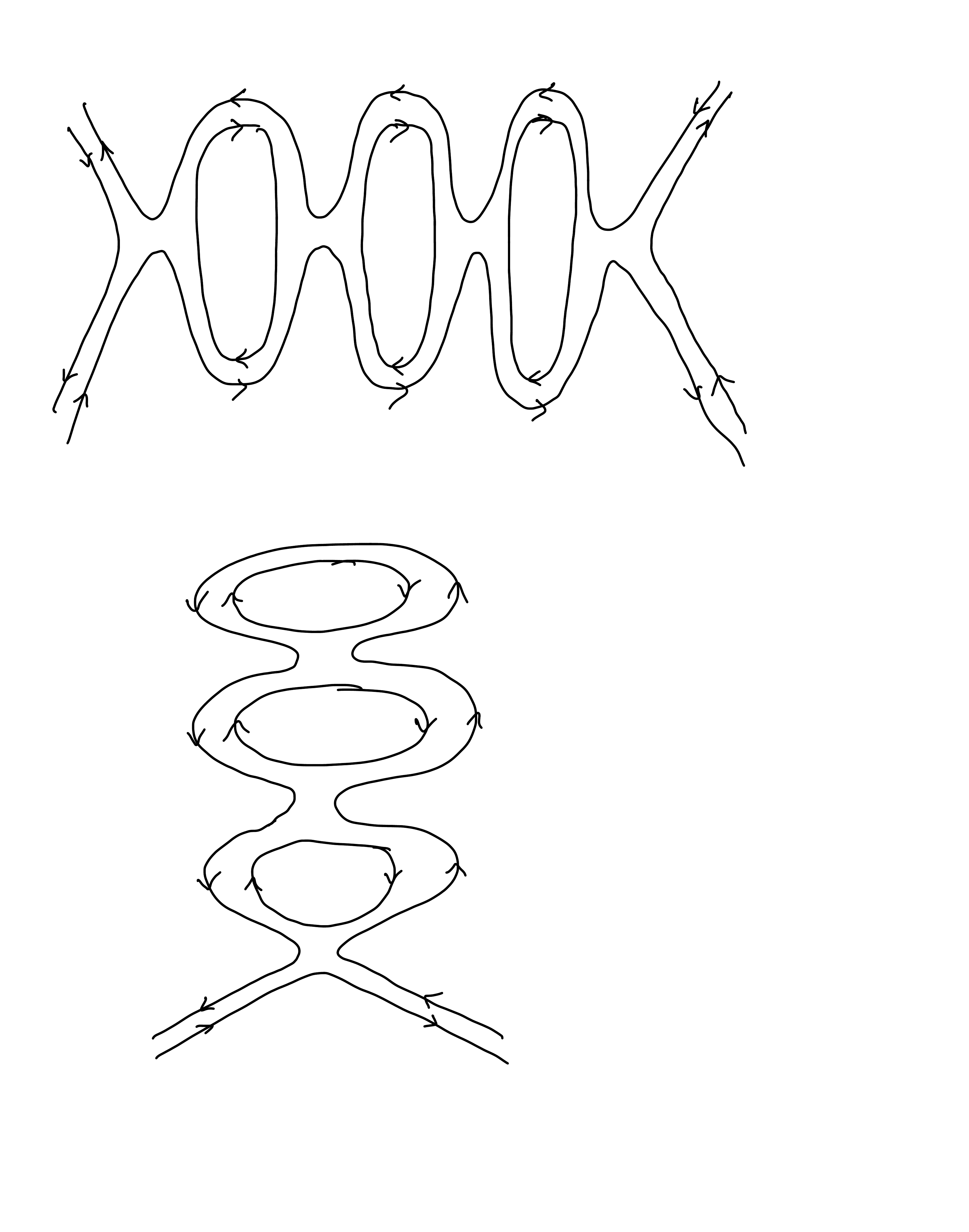}
 		\caption{A typical diagram that violates perturbative expansion in powers $\lambda_t$ in 't Hooft's double line notations. 
Each extra bubble brings an additional factor $\lambda_t$.  		 } 	
\label{bubbles}
 	\end{center}
 \end{figure}

 As the first step, let us have a closer look at the nature of would-be violation 
 of unitarity at strong 't Hooft coupling. The first place where this 
violation is manifest is the loop expansion.  An  example is given by bubble diagrams depicted on Fig.(\ref{bubbles}). 
Since the addition of each bubble  carries a factor  $\sim \lambda_t$, the expansion breaks down for 
large $\lambda_t$. From the first glance, one would think that such 
breakdown of unitarity is not fundamental and can be bypassed by re-summation.
While the bubble diagrams are resummable, the question is whether this procedure renders the saturation of unitarity unphysical.  We shall argue that this is not the case.  
 \\
 
The important processes to look at are the multi-particle amplitudes 
of the sort $2 \rightarrow  n$, in which the final $n$-particle state 
has the form (\ref{nstate}).  
We wish to show that such processes saturate unitarity whenever the inverse-coupling entropy bound (\ref{GiaC}) 
is saturated by the final state. The same applies to the area-law bound 
(\ref{GiaA}). 
This saturation 
is physical and cannot be removed by resummation.  \\

 However, in order to avoid confusion, we must keep a clear separation between the following two
 summations.  
 \begin{itemize}
  \item The first one is the resummation  
 of all Feynman diagrams that  contribute to the transition 
 amplitude into a specific  $n$-particle micro-state (\ref{nstate}).   

  \item  The second is the summation - in the {\it cross section} - over all micro-states (\ref{nstate}) that belong to the same classical macro-state.  
\end{itemize}
 
  We focus on the first one first.  
 
  \subsection{Exponential suppression of individual $n$-particle micro-states} 
     
 In order to clearly distinguish the false saturation of unitarity from the real one, 
 consider first a theory in which the final state particles 
 do not transform under any large symmetry group
 ${\mathcal G}$.  In this case, we can simply temporarily forget about  the label $j$ in the final state.  
  Of course, we still assume that the four-point coupling $\alpha$ is weak.
  In such a theory, we look for a transition from an initial $2$-particle state into a state  (\ref{nstate}). The latter contains a 
 high occupation number $n$ with some 
  characteristic momentum $q = 1/R$.   As already discussed, 
 the proper coherent superposition of such states
 (\ref{sol}) can be 
  viewed as a lump or a solitonic wave of a classical field.   \\
  
It is well-accepted (see, \cite{expOld1} -
 \cite{Monin}) 
 that the cross-section for such a process must be exponentially suppressed. 
This is true, despite the fact that the multiplicity of 
contributing Feynman diagrams grows factorially with $n$ already at the tree-level \cite{factorial},\cite{factorial2}. 
Namely, at large $n$ the  perturbative cross-section behaves as, 
 \begin{equation} \label{sigma} 
 \sigma_{2\rightarrow n} = c_n \, n! \alpha^{n} \,,  
 \end{equation}  
where only the leading factorial and exponential scalings in $n$ are 
displayed explicitly.
All the standard  
integration, not connected with the ${\mathcal G}$-degeneracy 
of the final state, is included in the prefactor 
$c_n$ which has proper dimensionality. 
In particular, if theory is gapless, $c_n$ will include 
the standard infrared dressing due to emission of infinitely-soft quanta. \\

 As explained previously, since the prefactor $c_n$ exhibits a power-law dependence on 
$n$, it is unimportant for physics close to saturation point 
 at large $n$.
Therefore, as previously, we set all such coefficients equal to one. 
 The maximal error we commit  with this 
setting is $\sim \frac{\ln(n)}{n}$.    
 \\

The factorial growth of the perturbative cross section 
(\ref{sigma}) creates a false impression
that at large $n$ unitarity can be saturated
(or even violated)  at weak coupling $\alpha$ 
by a single final micro-state. Or to put it differently,  
a classical object can saturate unitarity without summation 
over final states of internal degeneracy 
 ${\mathcal G}$.   This is not true, since for $n > \alpha^{-1}$ the growth of (\ref{sigma}) is unphysical and cannot be trusted.
The reason is that the perturbative expansion in 
$\alpha$ breaks down beyond this point. 
  \\

Indeed, thinking of cross section  in terms of expansion in series
 of $\alpha$, we must stop as soon as 
$\sigma_{2\rightarrow n}$ reaches the minimum in $n$. 
This happens at $n=\alpha^{-1}$, i.e., for the critical 
value of the collective coupling, 
\begin{equation}\label{OptimalT}
{\rm {\it Optimal~truncation:}}~\lambda_c =1 \, .
\end{equation}
Hence, we shall adopt this value of the collective coupling
as the point of 
{\it optimal truncation} of series in $\alpha$.  
It is highly instructive that this optimal value of $\lambda_c$ 
coincides with its critical value obtained by the self-sustainability
condition (\ref{balance}). This is no accident and it reveals how 
the information about the non-perturbative solitonic state 
 penetrates in the realm of scattering amplitudes.  \\

 Now, using Stirling approximation, 
it is easy to see that for the critical value (\ref{OptimalT}) 
the cross section (\ref{sigma}) is exponentially suppressed, 
\begin{equation} \label{noptimal} 
 \sigma_{2\rightarrow n} = {\rm e}^{-n} = {\rm e}^{-\frac{1}{\alpha}} \,. 
 \end{equation} 
 This suppression represents an embodiment of the difficulty of producing 
 a classical object in a two-particle scattering process. \\
 
 From (\ref{OptimalT}), it is clear that the expression  (\ref{sigma}) can only be trusted for $n \leqslant \alpha^{-1}$.  
 Beyond this point  it must be abandoned and 
 non-perturbative methods must be used.  This non-perturbative 
 analysis \cite{expOld1}-\cite{Son} confirm the exponential suppression of transitions to states with high occupation number $n$. \\
 
 However, for self-sufficiency,  in the appendix we present a refined version of a short-cut non-perturbative argument of \cite{Gia3}.
 It shows that for $n \gg \alpha^{-1}$ the cross-section of any 
 given $n$-particle state (\ref{nstate})  is suppressed 
 as 
  \begin{equation} \label{novercrit}
   \sigma_{2\rightarrow n} \lesssim
   n! n^{-n} \sim {\rm e}^{-n} \,.      
 \end{equation} 
Notice, this is only a consistency upper bound and in reality the suppression could be much stronger. 
However, the above upper bound is sufficient for our considerations.

\subsection{Entropy enhancement} 

 We thus adopt a physically justified picture that, in the absence 
 of large internal degeneracy ${\mathcal G}$, 
the cross section of producing a high-occupation number state  
is exponentially suppressed, as given by   
(\ref{noptimal}) and (\ref{novercrit}).  \\

However, in the presence of a large internal degeneracy
group ${\mathcal G}$, a new twist appears. 
The theory now can give rise to classical objects that saturate entropy bound (\ref{GiaC}).  From quantum field theory perspective they represent the high occupation number states with exponential degeneracy
$n_{\rm st} = {\rm e}^{\frac{1}{\alpha}}$. \\

In such a case, while the exponential suppression of the properly resummed individual processes (\ref{novercrit}) continues to hold, the number 
of processes 
that contribute into creation of a given classical object is exponentially large.   This number is equal to the number of micro-states $n_{\rm st}$
that belong to the same classical macro-state. 
The total cross section of production 
of the classical object is thus obtained by summing over all
such micro-states, 
\begin{equation} \label{Total} 
 \sigma = \sum_{\rm micr.st}^{n_{\rm st}} \sigma_{2\rightarrow n}\, . 
 \end{equation}
 Notice, here and below the notation $ \sigma$ refers 
 exclusively to the part of the cross-section that describes a creation of a
given classical object. \\

We are now ready to understand the fundamental  meaning of the 
inverse coupling bound (\ref{GiaC}) in terms of the unitarity of the scattering amplitudes. For this, let us first note that for large $n$ the summation 
over the micro-states in (\ref{Total}) reduces to a multiplication 
by the micro-state degeneracy factor $n_{\rm st} = {\rm e}^{S}$, 
\begin{equation} \label{TotalSn} 
 \sigma = \sigma_{2\rightarrow n} {\rm e}^{S}\, . 
 \end{equation}
Using (\ref{noptimal}), at the point of optimal truncation, $\lambda_c=1$, this becomes, 
 \begin{equation}\label{crossS} 
 \sigma \, = \, {\rm e}^{-\frac{1}{\alpha} + S} \, . 
 \end{equation}
From this expression it is clear 
 that the cross section (\ref{crossS}) saturates/violates unitarity whenever the entropy $S$ saturates/violates the bound (\ref{GiaC}). 
 That is, the number of micro-states $n_{\rm st}$ compensates the exponential 
suppression of individual amplitudes precisely when the classical object saturates 
the inverse-coupling entropy bound (\ref{GiaC}).  At this point $\sigma$
becomes an all-inclusive cross-section and the corresponding classical 
object becomes a {\it saturon}. \\

The above phenomenon comes from an additional enhancement 
of the cross section due to an internal degeneracy  ${\mathcal G}$.  This degeneracy 
is responsible for the 
maximal entropy of the classical final-state.  
This saturation cannot be removed by any resummation. 
As discussed above, this effect is very different from a  ``false" saturation of unitarity due to factorial multiplicity of 
Feynman diagrams of 
 individual amplitudes.  
 \\  
 
It is useful to translate the unitarity bound in terms of 't Hooft coupling.   
 For this, we again focus at the optimal truncation point 
 $n=\frac{1}{\alpha}$. 
Then, the individual cross sections are 
given by (\ref{noptimal}) and the total one is given by 
(\ref{crossS}). Expressing the entropy  $S$ through
(\ref{EntropyS}), we can rewrite (\ref{crossS}) as   
\begin{equation} \label{TotalOpt} 
 \sigma  \, = \,  
  \left(\left((1 + \lambda_t)\frac{1}{\rm e}\right)^\frac{1}{\lambda_t}
 (1 + \frac{1}{\lambda_t})\right )^N \,.      
 \end{equation}
The  critical value of $\lambda_t$ for which the above 
cross section saturates unitarity is,
\begin{equation}\label{Tsatur}
 {\it Unitarity~saturation}:~ \lambda_t  \simeq 0.54  \, .
\end{equation} 
Of course, $\lambda_t$ here must be understood as the {\it running}  
't Hooft coupling evaluated at the scale $q$. 
As it is clear from (\ref{Saturation}), the exact same value also saturates 
the entropy bound (\ref{GiaC}). \\

 We thus see that the cross section is saturated by a classical object 
 exactly when the latter saturates the inverse-coupling 
  entropy bound (\ref{GiaC}).   
The object therefore represents a saturon.  Its mass and the size are uniquely determined as, 
\begin{equation} \label{Msaturon}
 {\rm {\it Saturon~mass:}}~ M \sim \frac{q}{\alpha} \sim  \frac{1}{\alpha R} 
\end{equation} 
and 
\begin{equation} \label{Rsaturon}
 {\rm {\it Saturon~size:}}~ R \sim \frac{1}{q} \,, 
\end{equation} 
where $q$ is the scale at which the running 
't Hooft coupling reaches the critical value (\ref{Tsatur}). \\

It is clear that  simultaneously the area law bound (\ref{GiaA}) 
is also saturated. Indeed, the saturon state breaks spontaneously both the space translations as well as the internal symmetry
 that acts on index $j$. The decay constant of the resulting Goldstone 
 modes is $f = \frac{\sqrt{n}}q = \frac{q}{\sqrt{\alpha}}$. It is then obvious 
 from (\ref{Rsaturon}) that the final state entropy 
 $S=\frac{1}{\alpha}$ that saturates the inverse-coupling bound is 
 equal to the area of the saturon in units of the Goldstone decay constant $f$. \\ 
 
Finally, it is clear from (\ref{Msaturon}) and (\ref{Rsaturon}) that 
the Bekenstein  (\ref{Bek1})  bound is also saturated. 
 The saturon, therefore, saturates the combined bound (\ref{Gia}).  
 \\

     \begin{figure}
 	\begin{center}
        \includegraphics[width=0.53\textwidth]{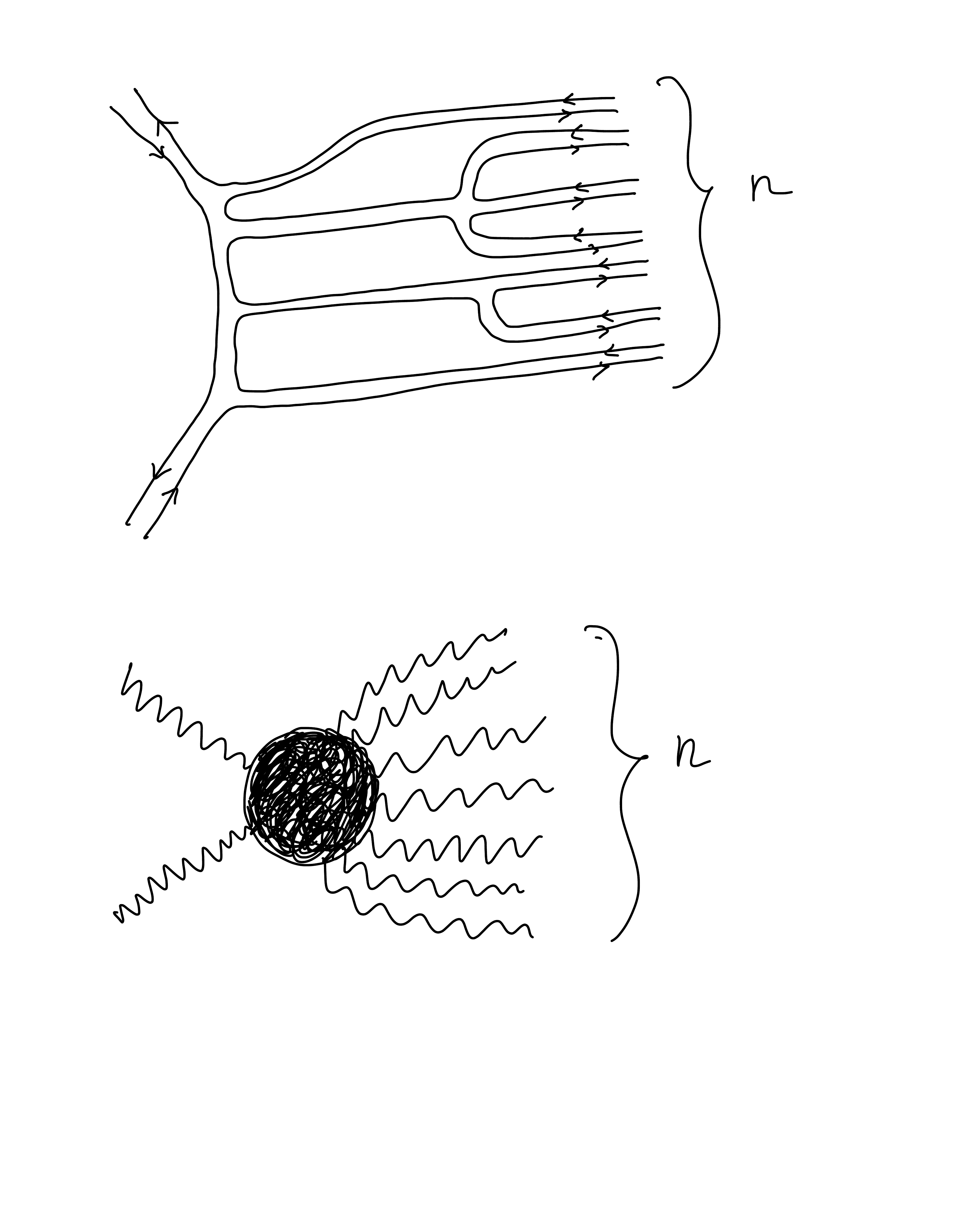}
 		\caption{A $2\rightarrow n$ process and an example of diagram in 't Hooft double-line notation contributing in it. } 	
\label{2toN}
 	\end{center}
 \end{figure}

The physical meaning of the above finding is pretty transparent. 
  When we form an $n$-particle state in a $2$-particle collision, we are effectively forming a classical object. 
   The formation probability is exponentially suppressed by 
   ${\rm e}^{-n}$. This suppression is confirmed both by 
  the previous analysis \cite{expOld1}-\cite{Son} as well as by the non-perturbative argument of \cite{Gia3} presented in the Appendix. \\
  
  However, when 
   the classical object saturates the entropy bound
  (\ref{GiaC}), the novelty appears.
   Now, the theory contains exponentially large number 
   of copies of the same classical object. I say ``copies" because 
   classically they are indistinguishable from one another. 
  Indeed, a classical observer, Alice, cannot resolve the  
  ``flavor" index $j$  since the coupling vanishes as 
  $\alpha \sim \frac{\lambda_t}{N} \sim \frac{\lambda_c}{n}$.
Rather, Alice is only sensitive to the  effects controlled by 
   't Hooft  and collective couplings.  
 That is,  Alice cannot tell the difference between the states 
 with different $j$-content, as long as the total occupation number 
 $n$ is large. \\
 
  Correspondingly, the production of any of these 
micro-states in a scattering experiment will be interpreted by Alice
as the production of {\it one and the same}  classical state.
Now, while each particular transition matrix element is exponentially 
suppressed, all of them will contribute to the Alice's
classical count.  Once the number of micro-states reaches the critical
value, this  classical object saturates the scattering cross-section. \\

    It is clear that this effect cannot be removed by any further re-summation.  Indeed, the resummation helps to compute the correct cross sections of the individual $2\rightarrow n$ physical processes. 
They come out exponentially suppressed (\ref{noptimal})-(\ref{novercrit}),  as they should.  At the same time, the  resummation cannot reduce the number of physically distinct 
 final states. As a result, no matter how suppressed are the individual 
 processes, the suppression gets compensated by the multiplicity of final 
 micro-states  when the corresponding micro-state entropy saturates 
the bound (\ref{GiaC}).   This is a fully non-perturbative phenomenon highlighting a deep connection between entropy and unitarity.

 \section{Entropic Meaning of Confinement }
  
  One remarkable thing in connection between entropy and unitarity  
  is that the saturation is fully controlled by
  't Hooft and collective couplings, $\lambda_t,\lambda_c$.  
 At the same time, the quantum coupling  $\alpha$
 can be arbitrarily weak. 
  It is fair to ask: \\
  
  {\it What happens if we try to deform the theory 
  and push the state beyond the saturation point? } \\
 
  This can be done by fixing the collective coupling at the critical 
  value  $\lambda_c = 1$ while increasing the 
  't Hooft coupling. 
  From  (\ref{Nstates}) it is clear that 
  for $\lambda_t \rightarrow \infty$ 
  the number  of micro-states increases as, 
  \begin{equation} \label{larget}
  n_{\rm st} \simeq ({\rm e} \lambda_t)^{\frac{1}{\alpha}} \,. 
  \end{equation}
  Correspondingly, the entropy of  the macro-state increases 
   as,
    \begin{equation} \label{larget}
  S \simeq \frac{1}{\alpha} \left(1+ \ln (\lambda_t)\right)\,.   
  \end{equation}
 Consequently, for $\lambda_t \gg 1$ the bound (\ref{GiaC}) is violated. 
 Simultaneously, the cross section (\ref{TotalOpt}) diverges as, 
 \begin{equation} \label{TotalOpt1} 
 \sigma \simeq (\lambda_t)^{\frac{1}{\alpha}} \, ,       
\end{equation}
and violates unitarity. Obviously, in a consistent theory this cannot happen.  What is the lesson that we are learning from here? \\

As a minimalistic move, we must adopt the saturation value as 
a {\it consistency upper bound}  
on 't Hooft coupling. The precise value depends on the 
representation content under the symmetry group ${\mathcal G}$ but 
in general is order one.   \\

 Yet, the story must be more profound.  
It would be somewhat counter-intuitive if a theory allows us 
to cross into a dangerous domain without a prior warning.  
Of course, one can say that violation of unitarity by a multi-particle state 
is a clear warning sign.  However, we expect that a consistent theory does not stop here. Instead, it must block the entrance into the dangerous domain of the parameter space {\it dynamically}.  
\\

 Therefore, we would like to ask whether a consistent theory possesses  a {\it built-in } mechanism that prevents such deformations from happening.  We shall now argue that confinement in
 $SU(N)$ gauge theory represents 
 such a preventive mechanism agains the violations of the entropy and 
 unitarity bounds. This idea has already been put forward in \cite{Gia1}
and we shall now elaborate on it.  

\subsection{Confinement from entropy bound}

As an illustrative example, we consider a $SU(N)$ Yang-Mills gauge theory 
with no fermions.  As it is well-known, this theory is asymptotically free, with  
the running gauge coupling $\alpha(q)$ becoming weak at short distances. 
We shall define  the 't Hooft coupling $\lambda_t$  as before 
(\ref{thooft}) and shall be working in 't Hooft's limit  (\ref{tlimit}).   
Obviously, in this limit QCD scale $\Lambda_{QCD}$ is kept fixed.  \\

 Now,  as shown in \cite{Gia2}, in this theory the entropy of 
an isolated instanton saturates the bounds 
(\ref{GiaA}) and (\ref{GiaC}) for a critical value of the 
't Hooft coupling $\lambda_t \sim 1$.  
For a generic value of $\lambda_t$, the entropy scaling is similar to (\ref{EntropyS}).
More details can be found 
in \cite{Gia2} and shall not be repeated here. 
Instead, we wish to establish what is the significance of this fact from the point of view of the scattering amplitudes. Next, we wish to find out 
 how the theory responds if we attempt to violate the bound 
by making $\lambda_t$ large. \\

First, we wish to show that the violations of the entropy bounds 
(\ref{GiaA})-(\ref{GiaC}) by instanton (or any colored state) would result into violation of unitarity by the scattering amplitudes.  We then 
 argue that this is prevented by confinement. 
We shall try to support this statement by assuming the opposite 
and running into an inconsistency. \\

 Indeed, assume that the theory never becomes confining. 
 Yet, it is asymptotically free and therefore is consistent in UV. 
 In such a theory there is no visible reason for 
 why we cannot force an  instanton of some size $R$ to violate the entropy bounds (\ref{GiaA}) and (\ref{GiaC}). This can always be achieved by making  the 't Hooft coupling $\lambda_t$ arbitrarily large at that scale. \\

However, the problem with this proposal is that simultaneously the unitarity would be violated 
 by a $2 \rightarrow n $ scattering process with the momentum-transfer 
 $q =\frac{1}{R}$. Consider a  process in which the two initial gluons would 
 scatter into $n$ final ones, 
 \begin{equation} \label{Atomany}
 A_{\beta}^{\gamma}A_{\gamma}^{\xi} \rightarrow 
 A_\beta^{\alpha_1}A_{\alpha_1}^{\alpha_2}A_{\alpha_2}^{\alpha_3}...
A_{\alpha_{n-1}}^{\xi}\, .
\end{equation}
A typical 't Hooft diagram describing a process of this  
sort is given in Fig.(\ref{2toN}). 
The color labels $\beta$ and $\xi$ are fixed by the initial gluons, whereas 
the color labels 
 $\alpha_{j}~(j=1,...,n-1)$ take values from $1$ to $N$. \\
 
 Since, by assumption, the theory is not confining, the complete set 
 of S-matrix asymptotic states can be represented by all possible $n$-gluon states with arbitrary color indexes,
 \begin{equation}  \label{completeset} 
 \ket{A_{\beta_1}^{\alpha_1}A_{\beta_2}^{\alpha_2}, ...,
A_{\beta_{n}}^{\alpha_n}}\,.
 \end{equation}
 Of course, by symmetry, the final state vector $\ket{t=\infty}$, obtained 
  as a result of Hamiltonian evolution, must transform under the same 
  representation of the $SU(N)$-group as the initial state 
  $\ket{t=-\infty}$.  That is, the state  $\ket{t=\infty}$ must transform as a hermitian
   traceless $N\times N$ matrix with respect to the open color 
 indexes $\xi$ and $\beta$.
  So, the true final state will be an appropriate superposition of all possible gluon states (\ref{completeset}). In the current example this superposition 
  will contain traces with respect to all indexes other than 
  $\xi$ and $\beta$. Schematically, 
   \begin{equation} \label{Traces}
 \ket{t=\infty} =\sum_n\sum_{\alpha_1,...,\alpha_{n-1}} u_n\ket{
 A_\beta^{\alpha_1}A_{\alpha_1}^{\alpha_2}A_{\alpha_2}^{\alpha_3}...
A_{\alpha_{n-1}}^{\xi}} \, , 
\end{equation}
where $u_n$ are some coefficients. 
The S-matrix elements will be determined by projecting this superposition
on different individual states from the complete set (\ref{completeset}).
Correspondingly, in the rate of the process the squares of S-matrix elements are  summed over all such states.  In particular, for $2\rightarrow n$ processes 
of the type (\ref{Atomany}) this amounts to,  
 \begin{equation} \label{sumA}
 \sum_{\alpha_1,...,\alpha_{n-1}} |\bra{A_{\beta}^{\gamma}A_{\gamma}^{\xi}}\hat{S}\ket{
 A_\beta^{\alpha_1}A_{\alpha_1}^{\alpha_2}A_{\alpha_2}^{\alpha_3}...
A_{\alpha_{n-1}}^{\xi}}|^2\, .
\end{equation}
In order to avoid a potential confusion with the counting of the final states,  we can softly Higgs the color group. We can easily achieve this 
by giving the tiny vacuum expectation values to a set of the ``spectator"  Higgs fields. 
 Such a  Higgsing of  $SU(N)$ symmetry generates a small mass gap
 and introduces the small mass splittings among the 
 gluon fields.   Since the theory is non-confining by assumption, 
 this splitting affects neither the structure nor the 
 magnitude of the amplitude.  However, it removes all doubts whether 
 the gluons of different colors must be counted as independent final states. 
  We can then smoothly take the vacuum expectation values of the Higgs fields to zero and recover a gapless theory. \\

 Note, in practice,  the assumption that we are in an unconfining theory
 means that the scale $R=q^{-1}$, at which the entropy bound is violated,   
  can be taken 
 {\it arbitrarily shorter}  than the length  of the confinement, $L_{QCD}$. 
For example, we can choose $L_{QCD}$ to be 
of galactic size, whereas $q=R^{-1}$ to correspond to LHC energies. 
Obviously, in such a case a local LHC observer is not affected by the 
confinement.  Such an observer would use the colored gluons 
(\ref{completeset}), rather than the colorless composites such as
glueballs,  as the asymptotic states of the S-matrix. Can 
such an observer witness a violation of entropy by some
field configuration at the scale $R$? 
 \\

In order to argue against this, first assume that we are dealing with a fully resummed amplitude. Then, our previous discussion goes through
and we skip the details.  
The summary is that the  cross section of creating an each particular $n$-gluon state is exponentially 
suppressed.  The enhancement is due to summation over 
micro-states corresponding to different color 
assignments of the final gluons, as expressed in (\ref{sumA}). 
 The resulting multiplicity factor is similar to 
(\ref{Nstates}).  So,   for $n=\frac{1}{\alpha}$ the cross section
is given by  (\ref{TotalOpt}). This cross-section saturates unitarity
for $\lambda_t \sim 1$.  This is strikingly close to a critical value for which, as observed in \cite{Gia2}, the entropy of a single instanton of the same scale saturates both bounds, (\ref{GiaA}) and (\ref{GiaC}). \\

 We now wish to see what happens if we try to violate these bounds
 by deforming the theory. 
    We can achieve this by freezing  
 $\lambda_c = 1$ while increasing  
 the 't Hooft coupling, $\lambda_t \rightarrow  \infty$. 
 Of course, as already 
 discussed, this would immediately result in 
 a non-perturbative violation of unitarity by the process
 (\ref{Atomany}) since the cross section grows  exponentially 
 (\ref{TotalOpt1}) with large $\lambda_t$.  
   However, our point is that the confinement will set in before this can 
   happen. \\

In other words,  as already noted, by taking the theory not be confining, 
we have implicitly assumed that the scale 
of confinement $L_{QCD}$ can be arbitrarily separated from the 
scale $R$ were the saturations of the entropy and unitarity bounds were taking place.  Or equivalently, $L_{QCD}$ can be arbitrarily larger 
than the saturon size $R$.  
 What theory tells us is that this was a {\it wrong} assumption.  \\

 We shall now explain why. 
Indeed,  the increase of  
 $\lambda_t$ at a fixed scale $q$ 
 represents a motion in the {\it space of theories}. This is because 
 we are changing the  relation between $\alpha(q)$ and $N$. 
  However, alternatively, we can view the same deformation as 
  a motion towards the IR-scale $q$ from  some UV-scale $q' >  q$ 
  within the {\it same theory}.  Since we keep $\lambda_c=1$, 
  this motion is accompanied by changing 
  the number $n$ of gluon constituents in the final state. 
  That is, within the same theory,
  we move  from one process at the UV scale $q'$ to a different process at the IR-scale $q$. \\
  
  If gluons were to remain the valid degrees of freedom down to arbitrarily 
  low energies,  such a descend towards IR could be continued indefinitely.  We would then sooner or later violate both entropy bounds
 (\ref{GiaA}) and (\ref{GiaC}).  Correspondingly,  the unitarity would 
also be violated.  This would mean that the $SU(N)$ gauge theory is inconsistent, despite being asymptotically free.  \\

Somehow, the theory must prevent this from happening. 
In a theory with pure glue, the only visible mechanism 
that can prevent such an unlimited descend towards IR is confinement.      
 That is, the theory must become confining before we manage 
  to make $\lambda_t$ sufficiently large and violate both entropy bounds and unitarity. 
Thus, in a large-$N$ theory of pure glue the confinement appears to be a direct 
consequence of the bounds (\ref{GiaA}) and (\ref{GiaC})
and of the unitarity constraints imposed by them.   
 \\

\subsection{Baryons} 

 Notice, we encounter a similar resistance if we try to violate 
 the entropy bound by quark bound-states.  As observed 
 in \cite{Gia1}, a baryon of large-$N$ QCD \cite{WittenN}
 saturates the entropy bound 
 when the number of quark flavors $N_F$  becomes of the same order as the number of colors $N$.  Indeed, consider a baryon 
 transforming as a symmetric tensor of rank $N$ under the flavor 
 group $SU(N_F)$.  Its entropy is given by \cite{Gia1},  
 \begin{equation} \label{baryonentropy} 
 S_{\rm bar} \simeq \frac{1}{\alpha} 
 \ln \left((1 + \frac{\lambda_c}{\lambda_F})^{\lambda_F}
 (1 + \frac{\lambda_F}{\lambda_c})^{\lambda_c} \right ) \,,      
\end{equation}
where we have defined the analog of the 't Hooft coupling 
with respect to the global $SU(N_F)$-flavor group, $\lambda_F \equiv \alpha N_F$.   The baryon consists of $N$ quarks
and has a size $R_{\rm bar} \sim  \Lambda_{QCD}^{-1}$. Therefore, the collective coupling evaluated at the scale $q= R_{\rm bar}^{-1}$ is 
$\lambda_c \sim 1$.  The above entropy then saturates the bound 
(\ref{GiaC}) for $\lambda_F \sim 1$. That is, the entropy  reaches 
the allowed maximum  for $N \sim N_F$. \\

 Simultaneously,  the area-law bound (\ref{GiaA}) 
as well as the Bekenstein bound (\ref{Bek1}) are also saturated.
Indeed, remembering that the pion decay constant is 
$f_{\pi} =\sqrt{N}\Lambda_{QCD}$ and the baryon mass 
is $M_{\rm bar} = N \Lambda_{QCD}$, we can write, 
\begin{equation}\label{baryonB}
 S_{\rm bar} = {1 \over \alpha} = {1 \over \alpha_{\pi}} 
 = (R_{\rm bar} f_{\pi})^2 =  M_{\rm bar} R_{\rm bar} \,, 
 \end{equation} 
  where $\alpha_{\pi} = \frac{q^2}{f_{\pi}^2}$ is the pion coupling 
  constant evaluated at the scale 
  $q= R_{\rm bar}^{-1} = \Lambda_{QCD} $. \\
  
  It is natural that at the same time the $2 \rightarrow N$ pion scattering cross section 
  saturates unitarity for the momentum-transfer set by the above  
  scale $q$.  This cross section is given by the expression 
  analogous to (\ref{TotalOpt}) with $\lambda_t$ substituted by 
 $\lambda_F$ and $\alpha$ by $\alpha_{\pi}$.   
 This process can be interpreted as the production of a classical lump
  of the pion field. More precisely, the final state can be viewed as an overlapping pair of the pion solitons, i.e., skyrmions \cite{skyrme}.  These solitons,  
as shown by Witten \cite{WittenS}, offer an effective description 
of the baryons at large $N$. 
  \\
 
 Now, we can try to violate the entropy bound by taking  
 $\lambda_F \gg 1$.   
   However, this is impossible because of the two reasons.
  First,  this would make the theory asymptotically 
 {\it not free}.  Simultaneously, the above multi-pion scattering process would violate unitarity at the scale $q \ll R_{\rm bar}^{-1}$.  This would 
 mean that the effective theory of pions breaks down at distances 
 much larger than the would-be size of a baryon.  So the latter 
 object cannot even be described within such a theory. Of course, the two
 responses are related. Namely, the low energy theory of pions 
 ``senses" that
 something is going wrong in the UV and responds to it via violations of unitarity by multi-pion amplitudes.   We thus observe that asymptotic freedom prevents 
 the violation of the entropy bounds. \\

 From the above point of view,  the conformal window 
 \cite{conformalwindow}  is of special interest.  Since the coupling is 
 at the fixed point, it appears that in such a regime the saturons 
 with the fixed number of constituents $n=N$ and arbitrarily large 
 sizes $R$ can exist. Correspondingly, their masses will assume 
 values (\ref{Msaturon}).  As a result, the entropy 
 of a saturon will be independent of its size and will be 
 fixed at the bound (\ref{Gia}).  In this respect, such saturons
would exhibit a scale-invariance.

\section{Scanning the Cross Section}

We now wish to scan the multi-particle cross section over different  
values of kinematic variables. 
For this, we need to parameterize 
$\sigma$ properly. 
First, we shall choose $n$ and $q$ as the scanning variables.   
Of course, in general, the number of active species $N$ can depend on the scale of
momentum-transfer $q$.  However, to start with, we assume  
$N$ to be independent of $q$.  
The scale-dependence of  the 't Hooft coupling $\lambda_t(q)$ then is uniquely determined by the running of $\alpha(q)$.  
Thus, the cross section effectively depends on two parameters
$(n,q)$, which can be traded for  $(\lambda_c, \lambda_t)$
 or $(E,q)$, and so on.  \\
 
 We shall perform the scanning in two different regimes. 
 In the first case, we scan $n$ for fixed $q$.  This is equivalent of scanning 
 over $\lambda_c$ and $E$ while keeping $\lambda_t$ and $\alpha$ fixed. 
 In the second case, we scan over $q$ (equivalently, over $\lambda_t$ and $E$) for the fixed values of $n$.

\subsection{Scanning $\lambda_c$}.

 We first freeze $\lambda_t$, $\alpha$ and the scale $q$ 
 by the saturation condition (\ref{Tsatur}) while allowing  $n$ (equivalently $\lambda_c$) to vary. 
 In this way, we scan over various processes in the same theory. These processes probe the same momentum transfer scale $q$
 but differ by the occupation number $n$ in the final state. 
 Obviously, they take place at different center of mass energies 
$E=nq$. \\

 Now, when we move $\lambda_c$ away from its critical point,  
 the resulting $n$-particle state
  saturates neither entropy bound nor unitarity. 
  In order to see this, let us write the total $n$-particle 
  cross section (\ref{TotalOpt}) 
  for generic values of $\lambda_c$ and $\lambda_t$
  \begin{equation} \label{TotalGeneral} 
 \sigma = \sigma_{2\rightarrow n}  
  \left((1 + \frac{\lambda_t}{\lambda_c})^{\lambda_c}
 (1 + \frac{\lambda_c}{\lambda_t})^{\lambda_t} \right )^ 
  {\frac{1}{\alpha}} \,,      
\end{equation}
where 
\begin{equation} 
  \sigma_{2\rightarrow n}  \lesssim
   \begin{cases}
     (\lambda_c^{-1}{\rm e})^{-\frac{\lambda_c}{\alpha}}  & \text{for} ~\lambda_c \leqslant 1, \\
     {\rm e}^{-\frac{\lambda_c}{\alpha}}
    & \text{for} ~\lambda_c >  1\, . 
\end{cases}
   \end{equation} 
  As previously, using the power of large-$N$, the non-exponential 
prefactor is set equal to one.     
 Of course, at the point of optimal truncation  
 $\lambda_c =1$ the equation (\ref{TotalGeneral}) 
 reproduces (\ref{TotalOpt}).  
 As already expressed by (\ref{Saturation})  and 
 (\ref{Tsatur}), 
 at this critical point both entropy and unitarity 
 are saturated  for 
 \begin{equation}\label{54}
 \lambda_t(q) \simeq 0.54\,. 
 \end{equation}
  Thus, if the running 't Hooft coupling 
reaches this critical value at some scale $q$,  the theory  gives rise to a saturon state in its spectrum. 
The mass and the radius of the saturon are given by  
(\ref{Msaturon}) 
 and  
(\ref{Rsaturon})
respectively. \\ 

The expression (\ref{TotalGeneral}) shows that the cross section 
of $n$-particle state is peaked 
at  $\lambda_c =1$ with the width $\sim \alpha$ 
and falls-off exponentially away from this point. 
In particular,   
  \begin{equation} \label{cases1}
   \sigma \lesssim
   \begin{cases}
    \lambda_t^{\frac{\lambda_c}{\alpha}}   & \text{for} ~\lambda_c \ll 1, \\   
  (\frac{{\rm e}\lambda_c}{\lambda_t})^{\frac{\lambda_t}{\alpha}}    
     {\rm e}^{-\frac{\lambda_c}{\alpha}}
    & \text{for} ~\lambda_c \gg 1\, . 
\end{cases}
   \end{equation} 
  This means that the saturon represents an exponentially narrow ``resonance" of the width $\sim \alpha$ in the spectrum of all possible $n$-particle states 
  of momentum $q$.  As explained above, the momentum 
  $q$ is defined by the criticality of the 't Hooft coupling
  (\ref{54}). \\

    Expressing $\lambda_c = \frac{E}{M}$
    in terms of the center of mass energy
   $E =nq$ and the saturon mass $M =\frac{q}{\alpha}$, 
    we can rewrite (\ref{cases1}) as, 
  \begin{equation} \label{Ecases1} 
   \sigma \lesssim
   \begin{cases}
   (\lambda_t)^{\frac{E}{M\alpha}}  & \text{for} ~ E \ll M, \\   
  (\frac{{\rm e}E}{\lambda_t M})^{\frac{\lambda_t}{\alpha}}    
     {\rm e}^{-\frac{E}{M\alpha}}
    & \text{for} ~ E \gg  M\, . 
\end{cases}
   \end{equation} 
Now, remembering that $\lambda_t \simeq 0.54$, 
it is clear that away from the resonance energy $E=M$ the cross section is exponentially suppressed. \\

   Thus,  for producing a saturon in a $2$-particle scattering experiment, 
 the center of mass energy must be fine tuned to the mass of the 
 saturon with an accuracy,   
 \begin{equation}\label{window1}
{\rm Saturation~window:}~~  
 {\Delta E \over M} \sim \alpha\, .
 \end{equation}

   This illustrates the price that one needs to pay for producing a classical object with an unsuppressed cross-section in a renormalizable theory. \\
   
   \subsection{Scanning $\lambda_t(q)$} 
   
  We now wish to scan the cross section over $q$ and $n$ while keeping 
 $\lambda_c =1$.  Then, the $q$-dependence enters the cross section
 (\ref{TotalOpt})  through the running  't Hooft coupling. 
 Taking the derivative of (\ref{TotalOpt}) 
  with respect to $q^2$, we get, 
 \begin{equation} \label{sigmarun2} 
 \frac{d}{dq^2} \ln(\sigma)  \simeq
- N  \ln\left ((1 + \lambda_t){\rm e}^{-1} \right)^{\frac{1}{\lambda_t}} 
\frac{d}{dq^2} \ln(\lambda_t)
  \,.      
 \end{equation}
 Around the saturation value (\ref{54}) this expression simplifies to, 
 \begin{equation} \label{sigmad1} 
 \frac{d}{dq^2} \ln(\sigma)  \simeq
 N \frac{d}{dq^2} \ln(\lambda_t)
  \,,        
 \end{equation}
 or equivalently, 
 \begin{equation} \label{sigmad2} 
 \frac{d\ln(\sigma)}{d\ln(\lambda_t)}  \simeq
 N  
  \,.       
 \end{equation}
 This result teaches us several things. 
First, around the saturation point the derivative of the cross section 
with respect to $\lambda_t$ scales as $N$. 
Thus,  the scale-dependence 
of the cross section is extremely sensitive to the scale-dependence 
of $\lambda_t$. \\

 Now, consider a theory that is asymptotically-free. In such a theory, 
 $\lambda_t$ runs with $q^2$ logarithmically.  Then,  assuming we are 
 not at the fixed point, we get,
  \begin{equation} \label{trun} 
 \frac{d\ln(\lambda_t)}{d\ln(q^2)}  \sim \lambda_t  \,.       
 \end{equation}
 Since, at the saturation point (\ref{54}) the r.h.s. of the above equation  is order one,
 the derivative is order one.  Then,  (\ref{sigmad1}) tells us that the 
 derivative of $\sigma$ around the same point is of order $N$. 
 Thus, 
 the cross section sharply diminishes as we move towards UV from
 the saturation point (\ref{54}). \\
 
 The motion towards infrared is more subtle.  Obviously, any further increase of 
 $\lambda_t$ is impossible without violating the entropy bound. 
 Thus, we see the following two possibilities: 
 \begin{itemize}
  \item  Either the theory 
hits an infrared fixed point; 
  \item  Or it develops a mass-gap.
\end{itemize}  
The latter can happen either due to confinement or a Higgs effect. 
In particular, as discussed above, in  $SU(N)$ gauge theory   
 without matter,  confinement appears to be the only mechanism 
 that can prevent the violation of the entropy bound.  \\

 It is impressive how profound the quantum field theory is. It tells us that there is no ``free-lunch" for producing 
 a classical object in a two-particle scattering experiment 
 at weak-coupling. This is true, despite the fact that 
   the object saturates the cross section at the right energy $E=M$.  
  The price is that the kinematic {\it window of opportunity}  is very narrow.  \\ 
   
  It is certainly remarkable that a classical object can be produced with an unsuppressed cross-section in a renormalizable  theory.   
 However, it dominates the cross-section only 
 for a particular ``resonant" value of the center of mass energy. 
 Away from it, the cross section falls off steeply. 
 Fundamentally, the following trade-off takes place. 
 The difficulty of producing a classical object in a quantum process manifests itself in an extremely precise choice of the center of mass energy in the scattering experiment.  \\
 
  Can saturons unitarize the cross section in a continuous range 
  of energies?  In renormalizable theories the difficulty is in maintaining the criticality relations such as 
 $\lambda_c \simeq 1, \lambda_t \simeq 0.54$ over a continuous range
 of scales.   
   In other words, the renormalizable theories do not possess 
   saturons of arbitrary masses and sizes unless theory is at 
 some non-trivial fixed point.  \\

   Here comes a profound difference with non-renormalizable theories such as gravity. 
   Gravity contains an almost continuous 
 spectrum of saturons  starting from the Planck mass and above.  
 These saturons are black holes.  
 This is the reason why  gravity can self-unitarize by black holes  
at arbitrarily high center of mass energies above the Planck mass. 
It is interesting to confront how violations of the entropy bound and unitarity are avoided by the two theories. 
 In $SU(N)$ gauge theory this is achieved by confinement which generates 
 a mass gap and  forces the asymptotic states to be colorless.  
In contrast, in gravity the entropy violation is avoided by offering a black hole 
for arbitrarily high energy. In this way, the entropy is kept at the saturation point for arbitrarily high  center of mass energy.

 \section{A model of Saturon as vacuum bubble}

  We shall now come up with an explicit renormalizable theory 
 that contains saturons.  
 This theory allows us to take different parameter choices 
  for which various entropy bounds are saturated by solitonic objects 
  of different sizes and energy.  We can then explicitly trace
  how the theory becomes inconsistent  if Bekenstein bound 
  (\ref{Bek1}) is obeyed without respecting the other two bounds
  (\ref{GiaA}) and (\ref{GiaC}).
  The conclusion is that a consistent theory must respect all three 
  bounds and saturate all three of them simultaneously  (\ref{Gia}). 
   \\

\subsection{The model} 

Consider a theory of a scalar field $\phi$ that transforms as an adjoint 
  representation of $SU(N)$ symmetry.  
  As usual,  the  latter can be written as $N\times N$ traceless hermitian matrix 
 $\phi_{\alpha}^{\beta}$, where $\alpha,\beta=1,2,...,N$. 
 In order not to blur 
   the effect by the confinement,  we shall keep the 
 $SU(N)$-symmetry global.  The 
 Lagrangian of the theory is, 
    \begin{equation} \label{LagN}
  L = \frac{1}{2} {\rm Tr} (\partial_{\mu} \phi \partial^{\mu} \phi) -
    V(\phi) \,, 
  \end{equation}
 where the scalar potential has the form, 
    \begin{equation} \label{adjoint}
    V(\phi) \, = \, \frac{\alpha}{2} {\rm Tr} \left ( f\phi  - (\phi^2 - \frac{I}{N}{\rm Tr}\phi^2 )\right )^2 \,.
  \end{equation}
Here, $I$ is the unit $N\times N$ matrix.  The vacuum equations, 
   \begin{equation} \label{vacua}
  f\phi_{\alpha}^{\beta}  - (\phi^2)_{\alpha}^{\beta} + \frac{\delta_{\alpha}^{\beta}}{N}{\rm Tr}\phi^2 \,  = \, 0\,,
  \end{equation}
have many degenerate solutions.  They correspond 
to spontaneous breaking of $SU(N)$ symmetry down to 
$SU(N-K)\times SU(K)\times U(1)$ subgroups for  values 
of $0< K < N$. In addition there exists an unbroken symmetry vacuum
with $\phi_{\alpha}^{\beta} =0$. \\

  All the above vacua  are equally good for our purposes. So, for definiteness, 
 we shall focus on the unbroken-symmetry vacuum $\phi=0$ and  
 the one with $K=1$. In the latter vacuum only the following component 
  \begin{equation} \label{COMP}
  \phi_{\alpha}^{\beta}  = \phi(x) \, {\rm diag} ((N-1), -1, ....,-1) 
  \frac{1}{\sqrt{N(N-1)}} \,,  
  \end{equation}
 has a non-zero expectation value.  Up to irrelevant $1/N$-corrections, 
 this  expectation value is equal to  
    \begin{equation} \label{VEV}
  \langle \phi \rangle  = f  \,. 
  \end{equation}
  Due to spontaneous breaking of global $SU(N)$
 symmetry down to $SU(N-1)\times U(1)$, 
 this vacuum 
  houses massless Goldstone species. Their number is,  
  \begin{equation}\label{NumberGold} 
  N_{\rm Gold} = 2(N-1) \,, 
 \end{equation} 
  and  their decay
 constants are given by $f$.  As usual, the 
 coupling ``constant" of these Goldstones, which we denote by 
 $\alpha_{\rm Gold}$, exhibits the following dependence 
 on the scale of momentum-transfer 
 $q$,  
 \begin{equation} \label{AlphaGold}
    \alpha_{\rm Gold} =  \frac{q^2}{f^2} \, .
 \end{equation} 
 Correspondingly, we define the 't Hooft coupling for Goldstones, 
 \begin{equation} \label{LambdaGold}
    \lambda_{\rm Gold} \equiv \alpha_{\rm Gold}\,  
 N_{\rm Gold} \simeq 2N\frac{q^2}{f^2} \, .
 \end{equation} 
  
 Since the vacuum (\ref{COMP}) is exactly degenerate with the one with unbroken symmetry, there exist domain walls that separate the two. 
The solution for a planar infinite wall has the form,  
 \begin{equation} \label{kinkN}
    \phi(x) \, = \, \frac{f}{2}\left (1 \pm  {\rm tanh}(\frac{xm}{2}) \right)\,,
  \end{equation}
where $x$ is a coordinate that is perpendicular to the wall. 
 The tension (energy per unit surface area) of the wall is given by, 
  \begin{equation} \label{tension}
    \mu  \, = \, \frac{1}{6}\frac{m^3}{\alpha}\,,
  \end{equation}
   and the thickness of the wall is, 
    \begin{equation} \label{thickness}
    R  \, \sim \, \frac{1}{m}\,.
  \end{equation}
  Approximately, the same expressions apply to a 
  closed bubble when its radius $r$ is much larger than the wall thickness, 
  $r \gg R \sim m^{-1}$. This regime is usually referred to as the 
  {\it thin wall}  approximation. 
  \\
  
  In the regime of our interest, in which $\alpha$ is very small, 
 the bubbles are long-lived. That is, they oscillate for a sufficiently long time before decaying into particles.  The qualitative way for understanding this stability is different for large and for small bubbles. 
 For large bubbles ($r \gg m^{-1}$) the oscillation frequency is $\sim 1/r$. This is much less that the mass of a free quantum.  Consequently, the production rate is suppressed.   
  The decay rate for  the small bubbles,  $r \sim m^{-1}$, 
  will be derived later. However, a qualitative reason for their
 long life-time is that 
  the decay goes through the quantum re-scattering of constituents 
 which is suppressed due to weak coupling. 
  \\  
   
 Notice, if we restrict the adjoint field to its component 
  (\ref{COMP}), the potential (\ref{adjoint}) becomes
     \begin{equation} \label{adjoint3}
    V(\phi) \, = \, \frac{\alpha}{2} \left ( f\phi  - \phi^2 \right )^2 \,  + \, 
  {\mathcal  O}(N^{-2}) \,. 
  \end{equation}  \\

 We now wish to derive the restrictions imposed on the theory by the 
 three entropy bounds,  (\ref{Bek1}), (\ref{GiaC}) and (\ref{GiaA})
 and by unitarity. 
  We start by choosing the trivial vacuum $\phi=0$ as our asymptotic 
  S-matrix vacuum.  In this vacuum  all particles have a mass 
  $m=\sqrt{\alpha}f$. Next, consider a vacuum bubble inside of which $\phi =f$.  \\

  The crucial fact 
  is that inside the bubble the $SU(N)$-symmetry is spontaneously 
  broken down to  $SU(N-1)\times U(1)$ subgroup. This breaking 
  results into $\sim 2N$ gapless Goldstone modes localized within the bubble world-volume.  These Goldstone modes create an exponentially large number of the bubble micro-states.  Using the method of  \cite{Gia1}, we can estimate this number in the following way.  The degeneracy of the 
 bubble interior is controlled by the degeneracy of the 
 vacuum manifold in the broken phase. This vacuum manifold is
 obtained by the action of  
 $SU(N)/SU(N-1)\times U(1)$ transformations on the 
 expectation value (\ref{COMP}). 
 The effective quantum Hamiltonian that describes the corresponding  
 degeneracy of the bubble is: 
 \begin{equation} \label{Hamilton}
 \hat{H} = X\left ( \sum_j \hat{a}_j^{\dagger}\hat{a}_j  -  s(r)\right ) \,, 
 \end{equation}
where $\hat{a}_j$-s are quantized zero modes that classically parameterize the bubble moduli space. Their number is of order $2N$.  The quantity $s(r)$ 
is the time-averaged space integral 
of $\phi^2(x)$.   For large (and slow) bubbles, $r\gg m$, for 
which the thin wall approximation works, it is given by the bubble 
volume times $mf^2$,
\begin{equation} \label{sfactorL} 
s(r) \, \simeq \, \frac{4\pi}{3} r^3mf^2 =  \frac{4\pi}{3} \frac{(rm)^3}{\alpha}\,,
\end{equation} 
whereas for the smallest bubbles, $r \sim m^{-1}$, we have  
$s \sim \frac{1}{\alpha}$. \\
  
   Now, the degeneracy of (\ref{Hamilton}) 
is given by the binomial factor which is of order 
 \begin{equation} \label{NstateBubble} 
n_{\rm st}(r) \sim   \left (1 + \frac{2N}{s(r)} \right )^{s(r)}
 \left (1 + \frac{s(r)}{2N} \right )^
 {2N} \,.   
\end{equation} 
This degeneracy endows the bubble with the corresponding micro-state entropy $S_{\rm bub}(r) = \ln(n_{\rm st}(r))$.
Next, for convenience, we introduce a notation,  
\begin{equation} \label{newlambda}
\lambda(r) \equiv \frac{2N}{s(r)} = \frac{2\lambda_t}{\alpha s(r)} \,, 
\end{equation}
where the 't Hooft coupling $\lambda_t$ is defined as before,
(\ref{thooft}). 
In this notations, we can 
write the entropy of a bubble of radius $r$ as  
\begin{equation} \label{EntropyBubble} 
S_{\rm bub}(r) = s(r) \ln\left (\left (1 + \lambda(r)
\right )
 \left (1 + \frac{1}{\lambda(r)}\right )^
 {\lambda(r)} \right )
\,.   
\end{equation} 
We shall now investigate the response of the theory 
when the above entropy saturates the three bounds
(\ref{Bek1}), (\ref{GiaA}) and (\ref{GiaC}) for the bubbles 
of various sizes.  

\subsubsection{Small bubbles as saturons}  
 
  We consider the smallest bubbles first, $r\sim R = m^{-1}$.
  The energy and the surface area of such a bubble are given 
  by $E_{\rm bub} \sim \frac{1}{R\alpha}$ and  Area$\sim R^2 \sim m^{-2}$ 
  respectively. 
 Correspondingly, for such bubbles we have, 
  \begin{equation} \label{SmallBubbleEq} 
  E_{\rm bub} R \sim \frac{1}{\alpha} \sim 
  \frac{1}{\alpha_{\rm Gold}} \,  
  \sim \, (Rf)^2 \, .
  \end{equation}
  Thus, all three bounds: The Bekenstein bound 
 (\ref{Bek1}), the inverse-coupling bound (\ref{GiaC}) and 
 the area-law bound (\ref{GiaA}) are satisfied simultaneously. 
 Moreover, the inverse-coupling bound is satisfied for both couplings: For the coupling of massless Goldstones, $\alpha_{\rm Gold}$, as well as, for the coupling of massive $\phi$-quanta, $\alpha$. The reason is that 
 the range of the  interactions mediated by both fields is large enough 
 to cover the size of the smallest bubble $r \sim m^{-1}$. 
 Correspondingly the bound (\ref{GiaC}) must be satisfied with respect to both couplings, and it is.  
 To put is shortly, we see that for smallest bubbles the relation (\ref{Gia}) holds.  \\

 From the definition (\ref{newlambda}) and the expression 
 (\ref{EntropyBubble}) it is easy to see that the above saturation 
 takes place when the both 't Hooft couplings are order one, 
 \begin{equation} \label{2thoofts}
 {\it Saturation~point:}~~\lambda_t \sim \lambda_{\rm Gold} \sim 1 \,.  
 \end{equation} 

Using our previous knowledge, it is easy to see how 
the above 
saturation of the entropy bound  is mapped on the saturation of unitarity. 
Namely, in respective S-matrix vacua  the processes $2\rightarrow n$
saturate unitarity at momentum transfer $q =m$. 
Of course, in both vacua, the saturation takes place at the points 
of optimal truncation. \\

  A typical process of this sort is 
 given by  Fig. (\ref{2toN}). Here the double lines must be understood as 
 the adjoint $\phi$-field in 't Hooft's notations.  For such processes, our 
 previous analysis is directly applicable.  As we already discussed in details, 
 the cross section of this process is given by 
 (\ref{TotalOpt}). Obviously, in this expression  we must insert the couplings  
that are relevant for a given process. For example, for
$2\rightarrow n$ Goldstone 
scattering process in $SU(N-1)\times U(1)$ vacuum, 
at the point of optimal truncation $n = \alpha_{\rm Gold}^{-1}$ 
the cross section 
will take the form, 
 \begin{equation} \label{CrossGoldOpt} 
 \sigma_{\rm Gold} =   
  \left({\rm e}^{-1}(1 + \lambda_{\rm Gold})
 (1 + \frac{1}{\lambda_{\rm Gold}})^{\lambda_{\rm Gold}} \right )^ 
  {\frac{1}{\alpha_{\rm Gold}}} \,.       
\end{equation} 
As we already discussed several times, the above cross section is 
saturated for $\lambda_{\rm Gold}$ order one. \\

It is not surprising that this matches a regime in which the 
 vacuum bubble saturates the entropy bound (\ref{Gia}). 
 Indeed, from the point of view of an S-matrix vacuum with unbroken 
 symmetry,  the smallest bubbles are well described as self-sustained states 
 of weakly interacting quanta of occupation number 
 $n = \frac{1}{\alpha}$. 
Correspondingly,  the $n$-particle process that saturates unitarity 
can be viewed as describing the formation of such a bubble in 
a two-particle scattering process.  \\

 As we have discussed previously, the processes with the higher number  of the final quanta are exponentially suppressed. The reason was that, 
 once we saturate the entropy bound by a state $\lambda_c =1$, all the states $\lambda_c \gg 1$ are well below the bound. 
 As a result, their entropy factors are too weak for winning over the exponential suppression of the amplitudes. 
 This is clearly illustrated by the equations (\ref{cases1}) and 
 (\ref{Ecases1}).  \\
 
 The above insufficiency of the entropy enhancement for the states 
 with $\lambda_c \gg 1$ is also matched by the entropy count of the larger bubbles $r \gg m^{-1}$.
  In order to see this, first check the entropies of
  such bubbles. 
  From (\ref{newlambda}) and 
 (\ref{sfactorL}) it is clear that for large bubbles we have
 \begin{equation} \label{lambdaLr} 
 \lambda(r) \simeq \frac{3\lambda_t}{2\pi(rm)^3}\,. 
 \end{equation} 
 Recall that the 't Hooft coupling was already 
set to its critical value $\lambda_t \sim 1$ by the requirement of entropy saturation by the smallest bubbles. Since, $\lambda_t$ is a parameter of theory, 
it is the same for the  bubbles of all sizes.  Then, 
from (\ref{EntropyBubble}) and  (\ref{lambdaLr})
it is clear that for the large size 
bubbles the entropy scales as, 
\begin{equation} \label{LargeBubbleL} 
S_{\rm bub}(r)|_{r \gg m^{-1}} \simeq \frac{2\lambda_t}{\alpha} 
\ln \left (\frac{2\pi {\rm e} (rm)^{3}}{3\lambda_t} \right )\,.   
\end{equation} 
It is not difficult to see that the above entropy is well below 
of all three bounds (\ref{Bek1}), 
(\ref{GiaA}) and  (\ref{GiaC}). \\

Indeed,  the maximal entropy permitted by the 
Bekenstein bound (\ref{Bek1}) for a large bubble has the form,
\begin{equation} \label{LargeBubbleBek} 
  S_{\rm Bek}(r) = 2\pi E_{\rm bub} r
  \simeq \frac{4\pi^2}{3} \frac{(rm)^{3}}{\alpha} \, .
  \end{equation}
 Obviously,  this is much larger than (\ref{LargeBubbleL}). \\
 
 Next, check the inverse coupling bound (\ref{GiaC}). Since the bubble 
 is much larger than $m^{-1}$, the only interaction that has a 
 relevant  
 range is the Goldstone exchange. Remembering that the Goldstone 
 coupling (\ref{AlphaGold}) evaluated at $q=\frac{1}{r}$ is 
 $\alpha_{\rm Gold} = (fr)^{-2}$, the  corresponding entropy bound 
 is 
 \begin{equation} \label{LargeBubbleAlpha} 
  S_{\rm Gold}(r) = \frac{1}{\alpha_{\rm Gold}} = 
   (fr)^2 = \frac{(mr)^2}{\alpha} \,.
 \end{equation} 
 Since the Goldstone coupling constant is equal to the inverse area
 of the bubble measured in units of $f$,
 the last expression also accounts for the area-law entropy bound
 (\ref{GiaA}). 
 As we can see, both are way larger than the actual entropy of the 
 large bubble (\ref{LargeBubbleL}).  \\  
  
To summarize, we see that when the smallest bubble saturates the 
entropy bound, it saturates all three bounds 
simultaneously, (\ref{Gia}). 
At the same time, the larger bubbles are below the bound.
Correspondingly, their entropies cannot compete against 
the exponential suppressions of the respective amplitudes. 
 \\

 \subsubsection{Suppression of large bubbles} 
 
  We  wish to explicitly demonstrate the insufficiency of the entropy enhancement of the cross sections for creation of  large bubbles in a two-particle scattering process in the regime in which the smallest bubbles saturate the entropy bound. We can achieve this by applying our results  to the analysis of \cite{bubbles}. 
 In this work a process 
 or bubble-creation in thin wall approximation was studied in a theory of a single real scalar field $\phi$ with two degenerate vacua. 
  Naturally, since such bubbles 
 carry zero entropy, no entropy enhancement was discussed there.   
  Notice, our theory would reduce to such a model if we would 
 reduce it to a single component (\ref{COMP}) of the adjoint field.
The resulting theory of course contains vacuum bubbles
 similar to ones we have studied.  However, they carry zero entropy due to the absence of the Goldstone phenomenon in the bubble interior. 
  Therefore, the bubble production rates in \cite{bubbles} and in 
 the present model (\ref{LagN}) will differ by the entropy factor.  
  \\

  In  \cite{bubbles} the creation of a vacuum 
  bubble of energy $E = nm$ from a single virtual quantum 
 was studied  
  as the first stage of a two-stage process. 
  The second stage amounts to a decay of the bubble into  
$n$ near-mass-threshold
  particles. 
  We shall focus on the first part of the process.  
In our notations, the matrix element of \cite{bubbles}
describing the bubble-formation has the form, 
 \begin{equation} \label{AmplitudeB}  
  |A_{1\rightarrow B}|^2  \sim {\rm e}^{-c n \sqrt{\lambda_c}} \,, 
  \end{equation} 
 where $c>0$.  \\

 Now, the novelty in our case is that the rate must be summed 
 over a large number of the bubble micro-states. This 
 amounts to multiplying 
  (\ref{AmplitudeB}) by the degeneracy factor 
 ${\rm e}^{S_{\rm bub}}$. For the large bubbles the entropy is given by 
 (\ref{LargeBubbleL}). Noticing that 
 $\lambda_c = \frac{E_{\rm bub}}{m}\alpha = \frac{2\pi}{3} (mr)^2$, 
 the large bubble entropy can be written as
 $S_{\rm bub}(r) \simeq \frac{3\lambda_t}{\alpha} 
\ln (\lambda_c )$.  The rate of the bubble production is then given by, 
  \begin{equation} \label{AmplitudeB1}  
 \Gamma \sim  |A_{1\rightarrow B}|^2 {\rm e}^{S_{\rm bub}} 
  \sim {\rm e}^{- n \sqrt{\lambda_c} \left(c - 3\lambda_t \frac{\ln (\lambda_c )}{\lambda_c\sqrt{\lambda_c}} \right )}\,. 
  \end{equation}  
Now, remembering that  in the above expression 
$\lambda_t \sim 1$ and $\lambda_c \gg 1$, it is clear that the entropy enhancement factor is negligible as compared to the suppression.  
So, the production rate of the large bubbles continues to be 
exponentially suppressed despite the entropy enhancement.  \\

Of course, the situation is very different for the smallest bubbles, 
$r \sim m^{-1}$, 
that saturate the entropy bound (\ref{Gia}). Because of this, they also saturate unitarity in the scattering process and are produced by an unsuppressed rate. This is also indicated by saturation of unitarity by 
the corresponding $n$-particle scattering process. \\

Unfortunately, the analysis of \cite{bubbles} is not applicable 
for small bubbles. Such bubbles correspond to $\lambda_c \sim 1$, 
which is outside of the validity domain of \cite{bubbles}.  
However, extrapolating (\ref{AmplitudeB1}) towards 
$\lambda_c \sim 1$, clearly shows the tendency:
The entropy factor starts to compensate the suppression term. 
Of course, this  is fully consistent with our results 
of saturating the $n$-particle cross section
at the optimal truncation point. 
This is natural since the smallest bubbles are well-described 
as $n$-particle states. Correspondingly,  the two pictures -  producing 
a bubble or an $n$-particle state - must match.  
 \\

\subsubsection{Superiority of area-law and inverse-coupling bounds}

 We now wish to show that saturating the Bekenstein bound 
(\ref{Bek1}) while disrespecting the bounds (\ref{GiaC}) and
(\ref{GiaA}) leads to an inconsistency of the theory.  
This indicates that in general the latter bounds are more stringent
than the former one. 
 \\

In the present model this happens when a large bubble 
 of certain radius $r_* \gg m^{-1}$ saturates the Bekenstein entropy bound (\ref{Bek1}).  As we shall see, such a saturation violates the other two bounds (\ref{GiaA}) and (\ref{GiaC}) and this triggers the  
violation of unitarity by the scattering amplitudes. \\ 

 The saturation value of $\lambda(r_*)$ can be found 
 by equating (\ref{EntropyBubble}) 
  to the corresponding 
 Bekenstein entropy (\ref{LargeBubbleBek}). Using the expression 
 (\ref{sfactorL}), this saturation condition  can be written in the following form,   
 \begin{equation} \label{SaturationLarge} 
\left (1 + \lambda(r_*)
\right )
 \left (1 + \frac{1}{\lambda(r_*)}\right )^
 {\lambda(r_*)}  \, \simeq \, {\rm e}^{\pi}  
\,,    
\end{equation} 
 which is satisfied for $\lambda(r_*) \simeq 8$. \\
 
 At first glance this saturation looks rather innocent. 
  However, meanwhile the bounds 
 (\ref{GiaA}) and (\ref{GiaC}) are violated both by the Goldstone coupling
 $\alpha_{\rm Gold}$ and the decay constant $f$.  
 This is immediately clear by comparing the 
maximal entropy (\ref{LargeBubbleAlpha}) permitted by 
the area (\ref{GiaA}) and the inverse-coupling (\ref{GiaC}) 
bounds to the Bekenstein entropy 
of the same bubble (\ref{LargeBubbleBek}). We have,  
 \begin{equation} \label{BekGold} 
  \frac{S_{\rm Bek}(r_*)}{S_{\rm Gold}(r_*)} = \frac{4\pi^2}{3} (mr_*) 
  \gg 1\,.
 \end{equation} 
The violation of the inverse-coupling (\ref{GiaC}) and the area-law (\ref{GiaA}) bounds, leads to the following disaster. \\

 First notice, that the corresponding value of the Goldstone 
 't Hooft coupling is enormously large, 
 \begin{equation} \label{lambdaGoldCrit} 
 \lambda_{\rm Gold} \simeq \frac{32\pi}{3}(r_*m)\, \gg \, 1.  
 \end{equation} 
 This is a very serious problem for the theory. 
  With such a strong 't Hooft coupling, the 
$2\rightarrow n$ Goldstone 
scattering process in $SU(N-1)\times U(1)$ vacuum,
violates unitarity at the point of optimal truncation $n = \alpha_{\rm Gold}^{-1}$.  Indeed, the  cross section (\ref{CrossGoldOpt}) for large Goldstone 't Hooft coupling given by   
(\ref{lambdaGoldCrit}) scales as, 
\begin{equation} \label{GoldViolationU} 
 \sigma \simeq (\lambda_{\rm Gold})^{\frac{1}{\alpha_{\rm Gold}}}  = 
 \left ( \frac{32\pi}{3}(r_*m)\right )^{\frac{(r_*m)^2}{\alpha}}\, .       
\end{equation}
 Since in this expression $(r_*m) \gg 1$, the above cross section violates 
 unitarity beyond any repair. \\
 
 Now, the important thing is that the above violation takes place 
 for the momentum-transfer $q \sim r_*^{-1}$. The
 physical meaning of this fact is that the actual UV-cutoff of the 
 theory in the Goldstone vacuum 
 is much less than the scale $r_*^{-1}$,
 \begin{equation} \label{UVR}
   \Lambda_{UV} \ll \frac{1}{r_*}\, .
  \end{equation}
 This means that the bubble of size $r_*$ cannot be described within 
 the validity of the theory. This is despite of the fact that the bubble 
 respects the standard Bekenstein bound (\ref{Bek1}).  
 It is the violation of the other two bounds (\ref{GiaA}) and (\ref{GiaC})  that makes the theory inconsistent. \\

 We thus arrive to the following conclusion: \\
 
 {\it  A violation of the inverse-coupling (\ref{GiaC}) and the area-law
 (\ref{GiaA}) entropy bounds makes the theory inconsistent
 even if the standard Bekenstein 
  bound (\ref{Bek1}) is satisfied.} \\

 Thus, the inverse-coupling (\ref{GiaC}) and the area-law
 (\ref{GiaA}) entropy bounds are not equivalent to Bekentein bound (\ref{Bek1}) and in fact are more 
 stringent. On the other hand, saturation of (\ref{GiaA}), (\ref{GiaC})
  also implies saturation of (\ref{Bek1}).
 Thus, in a consistent theory all three bounds must be respected 
 and saturated simultaneously.

  \section{Black Holes as Saturons}
  
 Obviously,  there are striking parallels exhibited by saturons in
 renormalizable theories on one hand and black holes in gravity on 
 the other.  These parallels
 appear to be so vast and so precise that they must indicate about the universality of physics-laws that govern the saturation point 
 (\ref{Gia}).
 This universality goes way beyond the particularities of the 
 underlying theory, whether it is gravity, a gauge theory
 or something entirely different. 
 What we are learning is that physics is controlled by a fundamental 
 connection between entropy and unitarity expressed by the bound
 (\ref{Gia}).  \\

 In this section we shall make these parallels 
 more transparent by organizing them in form of an explicit 
 ``checklist"  of similarities between {\it renormalizable}  saturons and black holes. 
 In order to make the extend of the connection brisk, we shall choose 
 for the role of non-gravitational saturons the vacuum bubbles of the theory 
 given by (\ref{LagN}).  We remind the reader that 
 the latter  is a {\it renormalizable} quantum field theory of a self-interacting scalar field $\phi$ in the adjoint representation of $SU(N)$ symmetry. Since this symmetry is not even gauged, it is hard to imagine an example that is more distant from gravity.  Nevertheless, as we shall 
 see, the saturons in this theory share all their key properties with black holes.  We shall now discuss these properties one by one. 

 \subsection{Similarities in entropy}   
 
 As  already discussed in details, saturons in the theory 
 (\ref{LagN}) represent vacuum bubbles.  An exterior of the bubble 
 is an unbroken symmetry vacuum which we choose as asymptotic S-matrix vacuum for our observer Alice. 
  In the interior of the bubble the 
 $SU(N)$ symmetry is spontaneously broken down to 
 a maximal subgroup which we chose as  $SU(N-1)\times U(1)$.
  This breaking results into $\sim N$ Goldstone bosons localized 
  in the bubble world volume.  They endow the bubble 
  with the entropy given by (\ref{EntropyS}).   As already explained, 
  the alternative way to think about bubble entropy is 
in terms of group representations. Because the bubble is not elementary 
but rather is a state with high occupation number, 
it transforms as a large representation 
  of the $SU(N)$ group. The entropy is set by the 
log of the dimensionality of this representation. 
As we have seen, only the smallest bubbles, of size $r \sim R=m^{-1}$,  
can saturate the entropy bound consistently. At the saturation point 
they saturate all three bounds (\ref{Bek1}), (\ref{GiaA}) and (\ref{GiaC})
simultaneously. Therefore, they  satisfy the relation (\ref{Gia}). \\

  Now, we wish to note that (\ref{Gia})  is exactly the relation satisfied by the Bekenstein entropy of a black hole 
 \cite{BekE}.  
Of course, the fact that black hole
entropy saturates  the ordinary Bekenstein bound (\ref{Bek1}) and   
 also exhibits the area law, is well-known. What is much less appreciated 
 is that the black hole entropy also saturates the 
 inverse-coupling bound (\ref{GiaC}). The latter observation was
 originally made in \cite{NP} which we shall now explain. \\
 
 For this, first note that the graviton coupling at the scale of momentum-transfer 
 $q$ is given by, 
 \begin{equation} \label{graviC} 
  \alpha_{\rm gr}(q) = \frac{q^2}{M_P^2} \, .
 \end{equation} 
  However, this is nothing but an inverse of the Bekenstein  entropy 
  of a black hole of radius $R = q^{-1}$! 
  Thus, the entropy of a black hole of mass $M$ and radius 
  $R = \frac{M}{M_P^2}$ 
  obeys the following relation,
  \begin{equation} \label{SBH}
  S_{\rm BH}  = MR = \frac{1}{\alpha_{\rm gr}(q)} = \frac{\rm Area}{M_P^{-2}}\,.       
 \end{equation} 
 This is exactly the relation (\ref{Gia}) 
 with $f= M_P$ and $q=1/R$.  
 As already explained in the introduction, the relation 
 is obvious since $M_P$ represents the graviton decay constant. 
 Also, a black hole breaks translation symmetry spontaneously 
 and the Goldstone mode of this breaking is of course the graviton excitation.

  \subsection{Decay and life-time} 
  
  Until now, the best understood computation about the 
 decay of a black hole,  is the famous original one by Hawking \cite{Hawking}. 
  This computation is exact in the following semi-classical limit, 
  \begin{equation} \label{HawkingLimit}
   M \rightarrow \infty, ~~ M_P \rightarrow \infty, ~~ R = {\rm finite} \,.  
   \end{equation} 
 Of course, simultaneously the Planck constant  $\hbar =1$ is kept finite.  
 Notice, in the above limit,  also the black hole entropy $S_{\rm BH}$ 
 becomes infinite, as it is clear from (\ref{SBH}). \\
 
 Now, in the limit (\ref{HawkingLimit}) the geometry of a black hole 
 experiences no back-reaction from the emitted quanta. That is, a black hole becomes a rigid reservoir of infinite energy 
 and information capacity.  The Hawking's computation shows that in this limit black hole emits 
 in thermal spectrum with temperature $T \sim \frac{1}{R}$. 
 That is, on average, a  black hole emits a quantum of energy 
 $\sim  \frac{1}{R}$ per time $\sim R$. The emission of more energetic 
 quanta is exponentially suppressed, whereas the less energetic ones are suppressed by the phase-space. \\

 Of course, in the limit (\ref{HawkingLimit}) the black hole mass is infinite and so is the life-time.  However, if we extrapolate Hawking's result 
 for finite $M$, we can estimate that the black hole shall lose 
 of order half of its mass approximately after the time, 
 \begin{equation} \label{lifeBH}
   t_{\rm BH}  \sim  R \, (RM_P)^2 \sim R \, S_{\rm BH} \, .
   \end{equation}   
 The last part of the equation relates this time-scale with the black hole entropy. This is indicative, since the number of the emitted quanta
 of energy $\sim 1/R$ is equal to the black hole entropy. \\
 
  Now, strictly speaking, it is unjustified to extrapolate the results of  Hawking's  semi-classical computation  beyond the above time-scale. The reason, 
  without entering into much guess-work  about the microscopic 
  quantum gravity, is simple \cite{GiaT}.  The back-reaction, that the black hole experiences with each emission, is of order $\sim \frac{1}{S_{\rm BH}}$.  So, the cumulative effect after the time  (\ref{lifeBH}) 
is large and must be taken into account.  This cannot be done without 
working in an explicit microscopic theory in which we shall not enter. 
  We shall therefore limit the study of the connection  between saturons
  and black holes by the time-scale (\ref{lifeBH}).  \\  
  
   We now wish to show that the {\it quantum}  decay of a saturon bubble exhibits a very similar behaviour.  Let us first note that the long life-time of {\it large}  bubbles 
  was concluded in the earlier studies both by numerics \cite{numerics} 
  as well as by analytic arguments \cite{bubbles}. The 
  latter argument relies  on a very narrow level-spacing of quantized bubbles.  
Due to this, the emission of particles requires transitions between distant levels
which is suppressed by the wave-function overlap. In the present case 
there will be an additional suppression factor due to the {\it memory  
burden} effect \cite{memory}. This effects is connected with the high entropy of the bubble which stabilizes it against the spread-out. 
Assuming that classically the bubble is long lived, we focus our interest on the smallest ones that saturate the 
entropy bounds and satisfy (\ref{Gia}).  \\

 Now,  for a saturon bubble of the theory
   (\ref{LagN}), the analog of Hawking's
   semi-classical limit (\ref{HawkingLimit}) is 
   \begin{equation} \label{SatLimit}
   M \rightarrow \infty, ~~ f \rightarrow \infty, ~~ R = {\rm finite} \,,  
   \end{equation}
   or equivalently, 
   \begin{equation} \label{SatLimit1}
   S_{\rm bubble} = \frac{1}{\alpha}  \rightarrow \infty,~~ 
   \lambda_c =1,~R = {\rm finite} \,.  
   \end{equation}
   In this limit, the decay rate of the saturon can be estimated in the following way.

   The saturon bubble represents a loose bound-state 
 of bosons of mass $m$. Because of 
 the binding potential their energies are of course below the 
 threshold of free quanta.  However,  the particles can be emitted because 
 of quantum depletion due to re-scattering.  The rate can be easily estimated and is given by (see, \cite{NP} for a very similar estimate of the depletion of a saturated state), 
  \begin{equation} \label{Satdecay}
  \Gamma_{\rm emission}  \sim R^{-1} \alpha^2n^2 \sim
  R^{-1} \, .
  \end{equation} 
  Thus, just like a black hole, the saturon emits on average one quantum of energy 
  $\sim \frac{1}{R}$ per time $\sim R$. 
  The emission of more energetic quanta is exponentially 
  suppressed because this requires a re-scattering of larger number 
  of constituents. 
    At the same time, the low energy ones are suppressed by the phase space. 
  Of course, since theory has a mass gap, nothing can be emitted 
  below the energy $\sim m$.  \\

  To summarize, an asymptotic 
  observer,  Alice, would see a saturon bubble as an object that emits in approximately-thermal spectrum. This is true despite the fact that the 
  $n$-particle state of saturon is not really thermal. What creates the 
  effect of thermality is the softness of the constituent quanta 
  and the fact that the state is at the critical point $\lambda_c=1$.  
  Now, extrapolating this result to finite $n$, the
 resulting half-life time of saturon bubble is, 
  \begin{equation} \label{lifeSat}
   t_{\rm bub}  \sim  R \, (Rf)^2 \sim R \, S_{\rm bub} \, .
   \end{equation}   
Without much commenting, the striking analogy with 
all the aspect of black hole evaporation and in particular with 
its half-life (\ref{lifeBH}) is obvious.  
 
  \subsection{Infomation horizon and time-scales}
  
  One of the characteristic properties of 
 semi-classical black holes (\ref{HawkingLimit}) is the existence 
 of the horizon. This makes an information about the black hole
 interior inaccessible for an outside observer, such as Alice. 
 It is widely believed, although remains a subject of active 
 controversy, that for a black hole of finite mass 
 the information is no longer hidden and finally comes out. 
  We shall not question this point of view since 
 within a consistent particle physics framework with unitary  S-matrix,
 no other outcome is imaginable for us. 
  The question therefore is not whether the information is accessible 
  but rather how long is the required time-scale for decoding it. 
 Of course, it is reasonable to assume that the  minimal 
 time-scale  required  for a start of the information read-out, is 
the half-life of a black hole. This view is supported by general
arguments by Page \cite{Page}. We shall therefore adopt 
the equation (\ref{lifeBH}) as the lower bound on such a 
time-scale.  \\

 We shall now see that all the above properties are matched by 
 saturons of renormalizable theory (\ref{LagN}). Of course, the advantage is that in case of a saturon bubble   
 we can understand the microscopic origin of such  
 properties very transparently.  Let us first notice that, just like a
 black hole,  a  saturon bubble 
 creates an {\it information horizon} that makes 
 the knowledge about its micro-state inaccessible for Alice.  
  Indeed, the quantum information is encoded in 
 saturon micro-states. These micro-states are labelled by 
 the excitations of the gapless Goldstone modes that are confined 
 to the interior of the saturon.  Their number is $\sim N$ as it is 
 also indicated by the entropy of the saturon. 
  \\
 
   Now, for reading out this information  Alice faces the following dilemma: 
  \begin{itemize}
  \item Alice can wait for Saturon evaporation and examine its decay 
 products very carefully;  
  \item Alternatively,  Alice can scatter an external probe particle at the saturon  and study the outcome. 
\end{itemize} 

 A slight technical problem with pursuing both methods simultaneously is that scattering will in general alter the internal state of saturon. So, it is cleaner to follow one protocol. \\

   It is easy to see that the minimal time-scale required by both efforts is   
 given by (\ref{lifeSat}). Indeed, in order to examine the 
 decay products carefully, Alice has to setup an interaction that 
 distinguishes  among the different states within the same 
 $SU(N)$-multiplet. 
 This is similar to measuring a spin polarization of a particle in a theory 
 with a rotationally-invariant  Hamiltonian. Despite the fact that 
 Hamiltonian commutes with the spin  operator, the particle
 spin projection can still be measured.  This is not an issue. 
  The problem in case of a saturon bubble  
is that the information is stored among the states of enormous 
number of Goldstone modes. So,  each emitted quantum carries 
only a tiny fraction of this information. \\

 The rate by which an emitted quantum 
interacts with Alice's device is, 
\begin{equation}\label{Alice1} 
  \Gamma_{\rm Alice} \sim \frac{1}{R} \alpha^2{\mathcal N}_{\rm Alice} \, ,  
\end{equation} 
where ${\mathcal N}_{\rm  Alice}$ is the measure of the capacity of the device which is under Alice's control.  Alice can maximize this capacity, 
for example, by preparing a huge reservoirs of probe particles. However, 
even if Alice manages to identify the state of a given emitted quantum, the latter only carries an exponentially small part of the 
information about the state of the entire saturon.   So,  Alice needs 
to gather at least of order $\sim  n$ emitted quanta 
before she can start decoding information at a reasonable rate.
This requires a minimal waiting time given by (\ref{lifeSat}), in 
exact analogy with a black hole.
 \\

Now, the second option for Alice is to scatter a {\it soft} probe particle through the 
interior of the saturon and study the scattering products. 
The hope is that the probe particle shall interact with Goldstone bosons 
that are confined within the interior and bring out the information about  
their state. Notice, the probe particle must be {\it optimally soft}:   
On one hand, it should not create too much level-splitting among the states of the gapless Goldstones and, on the other hand, the interaction rate 
must not be too low.  The latter rate is suppressed by the decay constant of Goldstone bosons $f$. \\

At the end, the rate of scattering 
between an optimally-soft probe and the saturon Goldstone field is, 
\begin{equation}\label{Alice2}
\Gamma_{Gold} \sim  \frac{1}{R^3f^2} \, .
\end{equation} 
The corresponding time-scale is nothing but a half-life of the saturon
bubble (\ref{lifeSat}).  Again, we observe that 
similarity with the black hole case is complete. In particular, in the limit 
(\ref{SatLimit}) the information becomes inaccessible. This is exactly analogous to what happens with black hole information in the limit 
(\ref{HawkingLimit}). Of course, both limits are  fully consistent with unitarity
since the respective objects become infinitely massive and 
their life-times become eternal.

\subsection{Scattering amplitudes} 
  
   As the last step for completing the list of similarities between 
  non-gravitational saturons and black holes, we discuss relation with scattering amplitudes.  As we have shown, the saturation of entropy bound 
  (\ref{Gia}) by a bubble (or any other soliton)  is in one to one correspondence with the sturation of  unitarity by the respective 
  $2\rightarrow n$ scattering process. 
  We wish to point out that this connections carries over into black holes. 
  The idea that a black hole can be produced in 
  a collision of few particles of center of mass energy $E \gg M_P$ 
  is not new and goes back to  
  \cite{BH1,BH2,BH3} and many subsequent papers. 
   However, only relatively recently \cite{Cl2},\cite{2N},\cite{2ND}, 
    this process has been connected to $2\rightarrow n$ graviton scattering amplitudes.  The actual detailed computation of the amplitude 
    was performed in \cite{2N} and \cite{2ND}.   
  The study was motivated by the microscopic picture of \cite{NP} 
 in which a black hole is described as $n$-graviton state 
 at the point of saturation $\lambda_c=1$.  However, in the present  discussion we would prefer not to have any microscopic bias. \\
  
  So, we put  ourselves in the position of  Alice, 
who is making no assumption about the microscopic theory of a black hole.   Alice is simply observing a process of black hole formation in a collision of 
two quanta of center of mass energy $E \gg M_P$ 
and its subsequent evaporation into $n$ soft ones. 
 It is obvious that the process that Alice identifies as a relevant 
 S-matrix process is $2\rightarrow n$.  \\

 This is exactly the computation performed in \cite{2N}.  The resulting  
cross section of producing  a particular $n$-graviton state is 
    \begin{equation} \label{2toBH}
   \sigma_{2\rightarrow n} \,  = \, 
   n! (\alpha_G)^{n} \,.        
 \end{equation} 
 The crucial point is that the above expression reduces 
 to,  
   \begin{equation} \label{2toBH1}
  \sigma_{2\rightarrow n}  = {\rm e}^{\frac{1}{\alpha_{\rm gr}}} =  {\rm e}^{-S_{\rm BH}} \,,       
 \end{equation} 
 exactly when the softness of outgoing gravitons matches 
 the  Hawking quanta $q = \frac{1}{R}$. Now, strictly speaking, we have 
 no moral obligation to interpret these  $n$-graviton states as the black hole micro-states. However, intuitively the connection is clear. 
 So, we can interpret them as ``relatives".  
 This relation carries the same meaning as the relation between 
 the saturon vacuum bubble in theory (\ref{LagN}) and the $n$-particle state into which it decays.  
 It is then clear that the total cross   
 section obtained by multiplying (\ref{2toBH1}) by the number of black hole micro-states, $n_{\rm st} = {\rm e}^{S_{\rm BH}}$, 
  saturates unitarity. \\
 
 In order to keep it sharp: In this discussion, we {\it do not} pretend 
 to understand the microscopic origin of $S_{\rm BH}$. Instead, we simply 
 take it for granted and observe that the structure of 
 the $2\rightarrow n$ graviton amplitude matches what is expected from a black hole. We are not going further than this.
  However, a complete similarity with the properties of a non-gravitational 
 saturon bubble -  
 where we do understand the microscopic origin of 
 the entropy - must ring some bell.  \\

  The above concludes our check list.  It is obvious from this 
  list that we are dealing with striking similarities between two types of
  objects.  On one side, this are saturons in a simple renormalizable theory. Their microscopic properties are as transparent 
  as they could be for a multi-particle state at weak coupling.  On the other side, we have black holes in a 
  non-renormalizable theory. Yet, we see that essentially all known properties match.  As we have seen, the central source that 
  defines these similarities 
  is that both saturate the bound (\ref{Gia}).  
  While the reader can decide for themselves how seriously to take this connection, our view is the following: \\
  
{\it  We think that there is something fundamental 
about the connection between saturations of unitarity and entropy 
encoded in the bounds (\ref{GiaA}) and (\ref{GiaC}).   
This connection goes well beyond gravity or renormalizability. 
 It is the saturation point (\ref{Gia}) that determines the behaviour
 of the system, including  its decay pattern, life-time, as well 
 as the capacities of information storage and processing. }

\section{Saturons and Classicalization}
  
  Few years ago \cite{Cl1} it has been suggested that 
 certain theories - that lack sensible Wilsonian UV-completions - can  
 instead be UV-completed by {\it classicalization}.
 The key idea is as follows. Consider a theory in which 
 a coupling $\alpha(q)$ becomes strong above certain cutoff  
 $\Lambda_{UV}$. In such a theory the processes with momentum-transfer $q > \Lambda_{UV}$ are out of control. 
 In certain cases the theory allows to be UV-completed 
above the scale $\Lambda_{UV}$ by integrating-in  
new weakly-interacting degrees of freedom. These new degrees of freedom restore perturbative unitarity in processes with 
momentum-transfer $q \gg  \Lambda_{UV}$. 
  We call such UV-completion {\it Wilsonian}. A nice example of this is
the Higgs in the Standard Model which restores unitarity in scattering of longitudinal $W$-bosons at high $q$. 
 What happens when the sensible Wilsonian UV-completion is 
 not possible? \\
 
  The idea of classicalization is that in such a case the theory can
use its classical objects for UV-completion.
A classical object of mass $M \gg \Lambda_{UV}$  and size 
$R \gg \Lambda_{UV}^{-1}$ 
is composed out of many soft quanta of momenta $q \sim  R^{-1}
\ll \Lambda_{UV}$.   Since $q$ is below the cutoff, the coupling 
is weak, $\alpha(q) \ll 1$. In this way, a would-be strong coupling 
is traded  for a high multiplicity.  
 Such a classical object represents a coherent state of the 
sort (\ref{sol}) with occupation 
number $n \sim \alpha(q)^{-1}$. 
\\

Now, imagine that a scattering 
process at center of mass energy $E \sim M \gg \Lambda_{UV}$  
is dominated by production of a classical state.  Such objects in 
\cite{Cl1} were referred to as {\it classicalons}.   
In such a case, the momentum-transfer in the process 
will be $q \ll \Lambda_{UV}$. This is because the constituent 
quanta of the classical objects are soft. 
Then, the process, despite being conducted at  
energy much higher than the cutoff,  never probes distances 
shorter than $R$.  So the theory shields itself from the strong 
coupling regime by  becoming {\it effectively-classical}. \\

However, there is a tradeoff: The occupation number 
must be very high.  Correspondingly, the theory must find a way 
of compensating the exponential suppression of the cross section
(\ref{novercrit}).  As explained in \cite{Gia3}, this
requires that the entropy of the classical object is high. 
Thus, in the language of present discussion, the classicalons must be saturons. 
 Then, from the results of the present paper it follows  
 that for classicalization to work,
the following two conditions must be satisfied: 
\begin{itemize}
  \item  The theory must contain saturons (classicalons);
  \item  Saturons must form an almost continuous spectrum for  
     $M > \Lambda_{UV}$. 
\end{itemize}
 The second requirement comes from our previous 
 findings that each saturon dominates the cross section only 
 in a very narrow window of center of mass energy 
 given by (\ref{window1}).  Therefore, a theory that is 
 UV-completed by classicalization must deliver a saturon 
for each value of the center of mass energy.  \\
 
 In a renormalizable asymptotically-free theory  the saturons appear
 with very specific masses (\ref{Msaturon}) and sizes 
 (\ref{Rsaturon}). These are  determined by the scale $q$ at which the 
 running 't Hooft coupling reaches the critical value (\ref{54}). 
 So, such a theory cannot be UV-completed by classicalization. 
 But, also there is no need for this since asymptotic-freedom 
 takes care of UV-physics.  \\
 
 On the other hand, non-renormalizable theories 
 can offer a continuous spectrum of saturons in UV. 
 The example of this is gravity.  There saturons are black holes.  
 This is why gravity can be unitarized by black hole creation.  
In fact, the proposal of UV-completion by classicalization \cite{Cl1} was based on a similar proposal 
for gravity \cite{SELF}. \\

Now, in order to avoid misunderstanding we must stress that 
unitarization by black holes works for center of mass energies above the   
Planck scale $M_P$.  In fact, higher the better. 
For processes with the center of mass energies  
$M_P$ in which  the momentum transfer is also 
of order $M_P$, the coupling $\alpha_{\rm gr}$ is order one. 
The resulting resonances produced in such collisions represent  
micro black holes.  These cannot be described classically. 
This is similar to production of  QCD resonances around
$\Lambda_{QCD}$ scale.  In the language of \cite{NP},  they 
are described as states with $n \sim 1$.

\section{Gravitational Species Bound} 

It has been shown \cite{species} 
that black hole physics 
puts the following bound on the number of particle species,  
\begin{equation} \label{Mstar}
 \Lambda_{UV}  \lesssim  {M_P \over \sqrt{N}} \, .
\end{equation} 
Here  $ \Lambda_{UV}$ represents the scale above which the quantum gravity 
enters the strong coupling regime to which the semi-classical treatment 
does not apply. \\
   
 Equation (\ref{Mstar}) is supported 
by several argument which can be found in \cite{species}
and will not be repeated here.  
 We just note that perhaps the physically most transparent one is 
 the following: 
A black hole of radius smaller than $\Lambda_{UV}^{-1}$ has 
no way  to sustain Hawking's thermal evaporation self-consistently.
Now, since Hawking's derivation is exact in semi-classical limit, 
 its invalidity implies a breakdown of semi-classical gravity. 
 Hence, the bound (\ref{Mstar}). \\

Because it relies exclusively on the validity of well-understood
properties of semi-classical black holes, the bound (\ref{Mstar}) is fully non-perturbative.
   The question therefore  is 
  whether this bound  can be understood in the language of scattering amplitudes. \\
  
\begin{figure}
 	\begin{center}
        \includegraphics[width=0.53\textwidth]{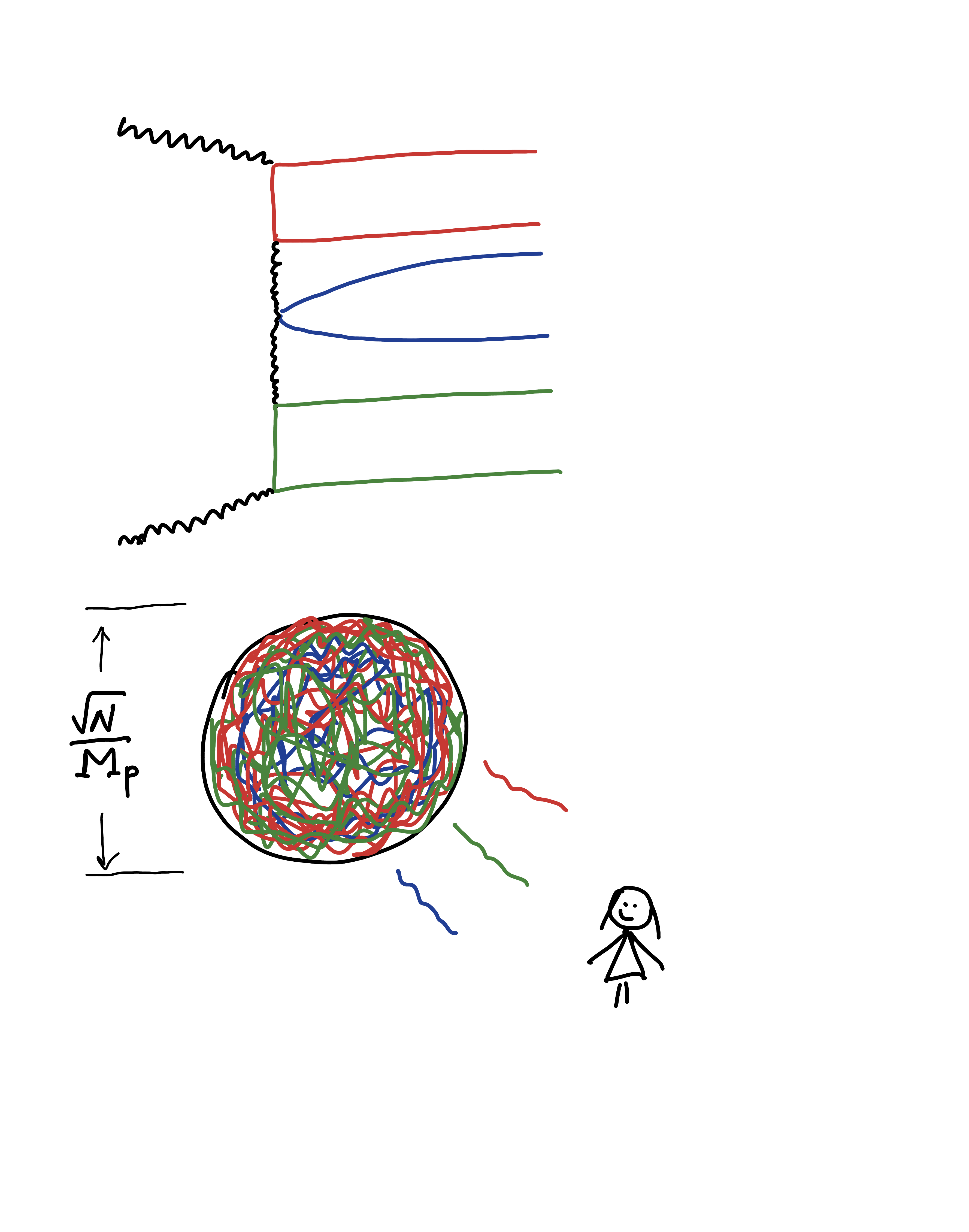}
 		\caption{Above, $2\rightarrow n$ process 
in which two initial gravitons produce $n$ particles
of different species denoted by different colors.
 The process saturates unitarity 
 at the species scale $\Lambda_{UV}  \sim  {M_P \over \sqrt{N}}$.
 Below, Alice observing a smallest semi-classical black hole which carries the species 
 hair (denoted by colors). }  	
\label{2Ngravity}
 	\end{center}
 \end{figure}

  The present discussion about the entropy saturation and unitarity 
 answers this question. 
 The relevant  processes 
 are the  $2\rightarrow n$ processes 
in which two initial gravitons produce $n$ particles of momenta  
$q \sim  {M_P \over \sqrt{N}}$. The example is depicted on Fig.(\ref{2Ngravity}).  Of course, the final state quanta gravitate and 
must be properly dresses by infrared gravitons.  This standard dressing is independent of entropy  
of species and is assumed to be done.   
Again, as before, by power of large-$N$ physics, all non-exponential 
and non-factorial dependences on $N$ play little role in determining 
the saturation point. Such factors therefore will be set to one. 
    \\
 
 Now, the  $n$  final-state particles can belong to $N$ different species
 and Einstein gravity couples to all of them democratically. 
 Due to this, the number of final states is exponentially large.   
The counting is identical to the one given for a gauge theory 
with a minor difference in a final degeneracy factor.  
 We shall display the cross section for
 \begin{equation} \label{gravityn}
n = {1 \over \alpha_{\rm gr}(q)} \,,  
\end{equation} 
 where, $\alpha_{\rm gr}(q)$  
  is the gravitational coupling given by (\ref{graviC}). 
 Defining the gravitational analog of the 't Hooft coupling, 
  \begin{equation} \label{gravityn}
 \lambda_{\rm gr} \equiv  \alpha_{\rm gr}N = \frac{q^2N}{M_P^2} \,, 
\end{equation} 
we can write the cross section in the form  
   \begin{equation} \label{TotalGravity} 
 \sigma =  
  \left({\rm e}^{-1}(1 + 2\lambda_{\rm gr})^{\frac{1}{2}}
 (1 + \frac{1}{2\lambda_{\rm gr}})^{\lambda_{\rm gr}} \right )^ 
  {\frac{1}{\alpha_{\rm gr}}} \,.       
\end{equation}
This cross section saturates unitarity for,   
\begin{equation} \label{11}
\lambda_{\rm gr} \simeq 1.1 \, .
\end{equation} 
Obviously, the corresponding value of momentum-transfer 
$q = \frac{M_P}{\sqrt{N}}$ marks the upper bound on UV-cutoff of the theory.  It is clear that this bound  is exactly the same 
as the species bound (\ref{Mstar}). \\

 We thus learn that the physical meaning of the species 
 scale $\Lambda_{UV}$ is the following. 
 It determines the  value of momentum-transfer $q$ that brings  
 the gravitational 
 't Hooft coupling to the saturation point (\ref{11}).  For this value, 
 the $n$-particle state becomes a saturon. That is, it saturates both the entropy bound  and unitarity.    
 This  saturon 
has a very clear physical meaning. It represents a  smallest 
possible semi-classical black hole. Such a black hole carries a species hair \cite{Nhair}. 
Notice that the entropy derived due to 
micro-state degeneracy of species, exactly matches the 
Bekenstein entropy of such a black hole.

\section{Outlook} 

In the present paper we have further explored the ideas 
about the connection between entropy and unitarity that were  
introduced in \cite{Gia1},\cite{Gia2}. 
The central message  is that unitarity of the scattering amplitudes 
imposes two universal bounds on the entropy of a quantum system. 
Namely, the maximal entropy is given by the area
measured in units of a decay constant $f$ of a relevant 
Goldstone degree of freedom (\ref{GiaA}). At the same time, 
the entropy bound is set  by the inverse running coupling 
$\alpha^{-1}$ 
evaluated at the scale of the size of the object (\ref{GiaC}). 
These bounds turn out to be more stringent and more general than 
Bekenstein's classic bound (\ref{Bek1}). In particular, they may be
violated by the objects that respect the latter bound. Of course, such systems are
eliminated by unitarity. 
 Also, 
since these bounds have no explicit reference to the energy, they
are applicable to the  Euclidean entities such as instantons for which the Bekenstein bound cannot be defined.  
On the other hand, the objects that saturate/respect  
(\ref{GiaA}) and (\ref{GiaC})  also saturate/respect the 
Bekenstein bound (\ref{Bek1}). That is, in a consistent theory 
all three bounds are saturated simultaneously. 
We refer to the objects that reach this point as {\it saturons}. \\ 

 We have seen that the saturation of both bounds (\ref{GiaA}) and (\ref{GiaC}) 
 is mapped on the saturation of unitarity 
 by $2 \rightarrow n$ scattering amplitudes with $n=\frac{1}{\alpha}$. 
 This saturation is non-perturbative.  Naturally, such processes are interpreted as the production of a saturon in two-particle collision. \\
 
Now, the saturon is a multi-particle state which is 
{\it approximately-classical}.  It therefore appears to defy  the standard 
field theoretic intuition that a production of a classical object  in a two-particle 
collision must be exponentially suppressed.  
 We have explained what is going on in reality. Fist, refining the analysis of \cite{Gia3},
 we gave a general argument 
 showing that the transition to each individual final state is indeed 
exponentially suppressed. This is in full accordance with the previous studies \cite{expOld1}-\cite{Monin}.  
 However, in case of a saturon the suppression is compensated by the exponentially large 
 number of micro-states that are 
{\it classically-indistinguishable}.  In other words, the cross section is 
enhanced due to the entropy of the final state. 
Due to this, with a properly chosen center of mass
energy, the saturon cross section can dominate the scattering process. 
However, the cross section is very narrowly  
 peaked at a resonant value of the initial energy.  Away from this value the cross section 
 diminishes exponentially steeply. \\
 
 Due to the above properties, saturons can play 
 the role in UV-completion by 
 classicalization \cite{Cl1}, but only if they form a continuous spectrum above certain energy. 
 However, it is unclear how wide is the range of such theories.  \\

 We have observed that consistent theories dynamically resist to violations of the entropy bounds.  
 An especially interesting example is provided by $SU(N)$ gauge theory. 
 It was already shown in \cite{Gia2} that an isolated instanton  
 saturates the entropy bounds (\ref{GiaA}) and (\ref{GiaC}) 
 at the critical value of 't Hooft coupling of order one.  
  We have seen that any further increase of the running  't Hooft coupling  
 would violate the entropy bounds. Correspondingly, the scattering amplitudes
 would violate unitarity. In order to prevent this from happening, the theory must become 
 confining. 
  This puts the phenomenon of confinement in a new
  light.  Namely,  it appears that in $SU(N)$ with pure glue 
  the  confinement  represents  a necessary response that avoids the violations of the entropy bounds and unitarity.  In other words, in order not to violate the entropy bounds (\ref{GiaA}) 
 and (\ref{GiaC}) somewhere in deep IR, the theory must eliminate the 
 asymptotic colored states. 
 The  possible alternatives would be that the theory either hits an IR fixed point 
 or develops a mass gap via the Higgs effect. However, none of the two options are feasible in pure glue.  
 Thus, confinement emerges as a direct consequence of the entropy bounds and unitarity. 
   \\
  
 Likewise, in \cite{Gia1}  it was observed that a baryon saturates the above entropy bounds when the numbers of flavors and colors are of the same order.  At this point the baryon entropy satisfies the relation
 (\ref{Gia}). The violation of the entropy bounds would render the theory 
 asymptotically non-free.  Simultaneously, the multi-pion scattering amplitudes would violate unitarity. \\

  Next, we have constructed an explicit theory that contains saturons. 
  We deliberately chose the example that is maximally distant 
  from gravity.  In particular, the theory is renormalizable and not
  based on any gauge symmetry.  The saturons  there represent 
 the vacuum bubbles that house a large number of Goldstone modes 
  in their interior.  These gapless Goldstone excitations create an exponentially large number of the bubble micro-states. 
    The resulting micro-state entropy
  saturates the bounds (\ref{GiaA}) and (\ref{GiaC})
 for a critical value of 't Hooft coupling. 
 At this point, the bubble becomes a saturon.  \\

   We have shown that on all counts the bubble saturons behave like black holes. 
   It is also clear that these properties are universal.  They must be shared by  saturons in other renomalizable theories. 
   The generalization of the constructions given in \cite{Gia1},\cite{Gia2} and in the  
present paper is straightforward.    
  In particular, for making contact with decaying black holes, we need to construct 
  saturons without any net conserved topological charge. 
   The vacuum bubble saturons  discussed in this paper 
  have this property.   
 The construction can  easily be generalized by creating saturons 
 using pairs of topological or non-topological solitons 
 with opposite charges that are placed on top of one another. 
 For example, one can pair up baryon-anti-baryon (skyrmion-anti-skyrmion), monopole-anti-monopole 
 and so on.  The annihilation of topological defect has been studied previously 
 numerically.  For example, monopole-anti-monopole pairs 
 were analysed in \cite{MMbar}.  However, to our knowledge, no studies have been done 
 either for the saturated case or in the limit (\ref{SatLimit1}). 
 The oscillating lumps of the scalar fields, the  so-called 
 oscillons \cite{oscillons}, can also be used as the building block
 for constructing a saturon.  However, one has to be careful 
 to stay within the regime of weak coupling $\alpha$. \\

 A profound  question for future studies is whether there 
are any implications of the present results for AdS/CFT correspondence \cite{ADS}.  Perhaps a natural avenue to go
would be to ask whether AdS can be viewed as a saturated state 
of some gravitational degrees of freedom, as it was suggested in \cite{NP}. No real progress in this direction 
has been achieved so far. Surprisingly, the analogous approach to 
de Sitter space turned out to be more straightforward.
In particular, the resolution of de Sitter patch in form of a saturated coherent state of gravitons  has been discussed in \cite{Qbreaking}. \\

    Finally, our studies bring us to the point at which the properties
  of a black hole can be understood through the prism 
  of a fundamental connection between unitarity and entropy. 
  We observe that this connection is universal and is shared by saturons irrespective 
  of their origin. This strongly suggest that black hole is
a saturon state of gravitons, as was originally proposed in \cite{NP}.  \\

\section*{Acknowledgements}
We thank Akaki Rusetsky and Goran Senjanovic for valuable discussions and comments. 
This work was supported in part by the Humboldt Foundation under Humboldt Professorship Award, by the Deutsche Forschungsgemeinschaft (DFG, German Research Foundation) under Germany's Excellence Strategy - EXC-2111 - 390814868,
and Germany's Excellence Strategy  under Excellence Cluster Origins.
 
 It is a great pleasure to thank Lorentz Institute and University of Leiden for 
 hospitality during the stay under the 2020 Lorentz Professorship.

 
 \section{Appendix: Argument from Effective $S$-Matrix}  
  We shall now give a fully non-perturbative consistency argument explaining why a properly resummed  matrix element 
   of transition
  \begin{equation}
  \ket{\rm few} \rightarrow \ket{\rm many}
  \end{equation}
  must be  exponentially suppressed. 
  This argument is a refined version of the one 
 in \cite{Gia3} and is based on effective $\hat{S}$-matrix.  
 Consider a process describing a transition between two sorts of quanta, denoted by $a$ and $b$ respectively.  During it,   
 $l$ particles of species $b$ get converted 
 into $n$ particles of species $a$.  Here, the term {\it species} 
 specifies all quantum numbers.  For, example 
 $a$ and $b$ can denote the different momentum modes of the same quantum field,  or some modes of two distinct fields. \\ 
  
 We assume that number eigenstates of $a$ and $b$ species 
 represent the legitimate 
 $S$-matrix states over the time-scales of interest.  
  Among other things, this implies that the effective Hamiltonian 
 is approximately diagonal in $a$ and $b$ modes throughout the 
 transition process. That is, the off-diagonal terms in the Hamiltonian 
 must be subleading as compared to the diagonal ones during the 
 relevant time-evolution.  This is a necessary condition for having a well-posed transition process.  It of course implies that the underlying field theory 
 stays within the weak-coupling regime throughout the transition. 
 The theory shall be otherwise unspecified.    
  \\

 We focus on the case when the occupation number $n$ in 
 the final state is much larger then the analogous number $l$ in the 
 initial state,  $\frac{n}{l} \gg 1$.   As we shall see, in such a case, the transition 
 matrix element is always exponentially suppressed. 
 This is in accordance with \cite{Son}. 
 Therefore, for simplicity we first take $l=1$.

 Thus,  the initial state is a one-particle state
 $\ket{in} = \ket{1}_b\otimes\ket{0}_a $ with a single 
 $b$-quantum 
 present.  Respectively, the final state $\ket{f} = \ket{0}_b\otimes\ket{n}_a $ is 
 populated with $n$  $a$-quanta.  Of course, we assume that the transition is kinematically allowed. \\
 
  Now, consider a fully resummed $\hat{S}$-matrix operator. 
 The  term that is responsible for the above transition has the form
 \begin{equation} \label{SM}
   \hat{S}_{1_b\rightarrow n_a} = \kappa (\hat{a}^{\dagger})^n \hat{b} \,.
 \end{equation} 
 The form is unique since the operator has to destroy a single particle 
 of species $b$ and create $n$ particles of species $a$. 
 Of course, the operator (\ref{SM})  is a result of resummation of infinite series. The information 
 about this resummation is contained in the coefficient $\kappa$.   
 We shall now argue that by consistency $\kappa$ is bounded as,   
 \begin{equation} \label{kappa}
   \kappa \lesssim  n^{-\frac{n}{2}}\,.   
 \end{equation} 
  This upper bound is universal and independent of the details of underlying field theory.  
It may come as a surprise because, naively, all we need to require is  
 that the matrix element satisfies, 
  \begin{equation} \label{naive1}
   |\bra{f}\hat{S}_{1_b\rightarrow n_a}\ket{in}|^2 < 1\,.    
 \end{equation} 
The latter requirement would give a much milder bound, 
 \begin{equation} \label{kappa2}
   \kappa <  \frac{1}{\sqrt{n!}}\,.   
 \end{equation} 
However, the correct bound is (\ref{kappa}). 
Here is why:  
In order to have 
a well-posed scattering problem, we must demand  
that the matrix element  $\bra{\psi}\hat{S}_{b\rightarrow na}\ket{\psi}$
is small over {\it all } the states $\ket{\psi}$ that are  
{\it physically  close} to either $\ket{in}$ or 
$\ket{f}$.  The meaning of this requirement we shall now explain. \\

We define the two normalized states $\ket{1}$ and $\ket{2}$  
as {\it physically close} if they provide comparable expectation values 
for a physical observable $\hat{O}$,  
\begin{equation}\label{closeO}
\bra{1}\hat{O}\ket{1} \sim \bra{2}\hat{O}\ket{2} \,. 
\end{equation}
Under {\it comparable}  we mean the same order of magnitude. 
The role of the physical observable $\hat{O}$ can be played by 
an arbitrary measurable quantity.   
We choose it to be 
 the number operator of $a$-quanta 
 $\hat{n} \equiv \hat{a}^{\dagger}\hat{a}$.  
 The reader should feel free to explore other choices.  
Then, according to above definition, 
a state $\ket{\psi}$ 
is {\it physically close}  if, for example, 
$\bra{\psi}\hat{n}\ket{\psi} \sim \bra{f}\hat{n}\ket{f}$. 
 Our criterion is that on any such state $\ket{\psi}$ the expectation value 
 of  $\hat{S}_{b\rightarrow na}$ must be small. 
 Why?  \\
 
 Here is one way to explain this. Think of the above transition process in  
terms of time-evolution in the 
 Hilbert space. Let the state vector at some initial time be 
 $\ket{t=0} = \ket{in}$. After a sufficiently long time $t$ this state evolves 
 into $\ket{t}$. The projection of $\bra{f}\ket{t=\infty}$ determines the 
 $S$-matrix elements. During the time evolution in any given process 
 the state vector explores only a finite portion of the infinite Hilbert space. 
 With the states populating this portion, vector $\ket{t}$ has a significant overlap. 
 These are states that are {\it physically-close} to $\ket{t}$. 
 Our requirement then is equivalent to demanding that 
 on all such states the off-diagonal part of the effective Hamiltonian 
 must be smaller than the diagonal part. A violation of this requirement 
 would imply that somewhere in the transition process 
 $a$ and $b$-modes stop to be the valid weakly-coupled
 degrees of freedom. The 
 Hamiltonian then must be re-diagonalized by a large canonical 
 transformation. This would contradict to our starting point.  \\

Since the state $\ket{\psi}$ can be chosen arbitrarily,  
we take it to be the following coherent state,
\begin{equation}\label{psi1}
 \ket{\psi} = {\rm e}^{\sqrt{n} (\hat{a}^{\dagger}
 - \hat{a})+ (\hat{b}^{\dagger} - \hat{b})} \ket{0} \,.
\end{equation}
Obviously, this state satisfies the criterion of the physical 
closeness since
\begin{equation} \label{close1}
\bra{\psi}\hat{n}\ket{\psi} = n = \bra{f}\hat{n}\ket{f} \,.
\end{equation} 
  Therefore,  we must require,  
  \begin{equation} \label{Ssmall}
  |\bra{\psi}\hat{S}_{1_b\rightarrow n_a}\ket{\psi}|^2 < 1\,,
  \end{equation} 
which  immediately gives (\ref{kappa}).   Taking this into account, we  
  get 
   \begin{equation} \label{Sn}
   |\bra{f}\hat{S}_{1_b\rightarrow n_a}\ket{in}|^2 < 
   n! n^{-n} \sim {\rm e}^{-n} \,,     
 \end{equation} 
 where in the last step we used Stirling's  approximation. 
 Thus, a transition matrix element, describing
 the creation of any $n$-particle state $\ket{f}$ from a one-particle initial state $\ket{in}$,  must be exponentially suppressed.  This is a non-perturbative result.  This conclusion is of course in  full agreement with the previous studies \cite{expOld1}-\cite{Monin}. However, it makes the origin of the suppression transparent from very general perspective 
 of $S$-matrix consistency.  \\
 
Obviously, the above reasoning can be easily generalized to the 
case in which the occupation number of $b$-particles in the 
initial state  
$\ket{in} = \ket{l}_b\otimes\ket{0}_a $ is larger than one.   As long as the 
 difference between the  occupation numbers
 in initial and final states 
is large,  $n \gg l$,   
 the exponential suppression of the transition matrix element takes place. 
 \\
 
  We shall now move to the case in which  the final particles
can belong to several different species.  That is,  we allow the operators     
 $\hat{a}_j$ to carry 
 a species label $j = 1,2,..,N$. This label can represent an arbitrary 
 quantum number such as ``color" or ``flavor".   
Thus, we are looking for a transition matrix element between an initial 
state $\ket{in} = \ket{1}_b\otimes\ket{0}_a $
and a final state $\ket{f} = \ket{0}_b\otimes \ket{n_1,n_2,...n_N}_a$, 
where  $\ket{n_1,n_2,...n_N}_a = 
\prod_{j=1}^{N} {(\hat{a}_j^{\dagger})^{n_j} \over \sqrt{n_j!}}
 \ket{0}_a $, with $\sum_{j=1}^N n_j =n$. The occupation  numbers $n_j$ are otherwise unconstrained. That is, the final state 
 $\ket{f}$ houses $n$-quanta with arbitrary color indexes. Of course, when only one color is occupied, 
 the story reduces to the case of singe $a$-species.  \\

  Correspondingly, the  transition $\hat{S}$-operator now has a form,  
  \begin{equation} \label{SMJ}
   \hat{S}_{b\rightarrow na} = 
   \kappa \prod_{j=1}^{N} (\hat{a}_j^{\dagger})^{n_j} \hat{b} \,, 
 \end{equation} 
  with the constraint $\sum_{j=1}^N n_j =n$.   \\
   
    In order to derive an upper bound on the coefficient $\kappa$, we shall repeat the previous reasoning. Namely, we demand a relative smallness of the expectation values of $\hat{S}$ over all the states 
$\ket{\psi}$  that 
 are {\it physically close} to $\ket{f}$. 
 Again, as a test observable we use the total number operator 
 of $a$-species,
 $\hat{n} \equiv \sum_{j=1}^N \hat{a}_j^{\dagger}\hat{a}_j$. 
 Correspondingly,  for 
 $\ket{\psi}$, we use 
a simple generalization of the state (\ref{psi1}) to several species, 
 \begin{equation}\label{psi}
 \ket{\psi} = {\rm e}^{\sum_{j=1}^N \sqrt{\tilde{n}_j} (\hat{a}_j^{\dagger}
 - \hat{a}_j)} \ket{0} \,.
\end{equation} 
  Here, we have introduced a notation tilde in order to distinguish 
    between the coherent state parameters $\tilde{n}_j$ 
   and the corresponding number eigenvalues $n_j$.
   We shall take $\tilde{n}_j \sim n_j$. Then,  
  \begin{equation} \label{close2}
\bra{\psi}\hat{n}\ket{\psi} = \sum_j \tilde{n}_j  \sim n 
= \bra{f}\hat{n}\ket{f} \,, 
\end{equation} 
which ensures that the states $\ket{\psi}$ and  $\ket{f}$ are physically close.  
    \\
     
Now, demanding the smallness of the 
expectation value (\ref{Ssmall}) evaluated for the 
$\hat{S}$-matrix operator (\ref{SMJ}) over the coherent state  (\ref{psi}),  
 we conclude that the coefficient $\kappa$ must obey, 
   \begin{equation} \label{kappa1}
   \kappa <  \prod_{j =1}^N \tilde{n}_j^{-\frac{n_j}{2}}\,.   
 \end{equation} 
  We shall now consider the cases of large and small values of 
  $n_j$ separately. We must remember that $n_j$-s are 
 characteristics of the transition process, whereas $\tilde{n}_j$ are parameters of the 
  probe state $\ket{\psi}$. The latter can be chosen at our convenience 
 subject to   
  $\tilde{n}_j \sim n_j$. 
   \\

 Now, for the case of large $n_j$-s, we can simply take $\tilde{n}_j = n_j$ 
 and use Stirling approximation in (\ref{kappa1}).
 Then, for the transition matrix element we get,
 \begin{equation} \label{Stran}
   |\bra{f}\hat{S}_{1_b\rightarrow n_a}\ket{in}|^2 < 
   \prod_j n_j! n_j^{-n_j} \sim {\rm e}^{-\sum_j n_j} 
=   {\rm e}^{-n} \,.    
 \end{equation} 
 
Regarding the case of $n_j \sim 1$, 
it suffices to 
take $\tilde{n}_j$ slightly larger than $n_j$. For example, 
consider the case $n_j=1$ for all $j$. Of course, in this case $n=N$.  
Taking $\tilde{n}_j = e n_j =e$, we see from 
(\ref{kappa1}) that
the transition matrix element is suppressed as 
${\rm e}^{-n} = {\rm e}^{-N}$.  \\

  In summary, we arrive to the universal suppression 
of a transition matrix element, 
\begin{equation} \label{Hiper}
|\bra{\rm many} \hat{S} \ket{\rm few}|^2 \lesssim e^{-(many)} \, . 
 \end{equation} 
 Here many=$n$ denotes the total occupation number in the final state.  
This result fully matches the physical intuition
which tells us that the creation of classical states in collisions of few quanta must be strongly suppressed. 
Indeed,  the transition $\ket{\rm few} \rightarrow \ket{\rm many}$ represents 
a quantum-to-classical transition.  The classicality of the final state 
is obvious when 
the occupation  numbers of the individual species, $n_j$,   
are large.  However,  the same is also true when the individual 
 numbers $n_j$ are small,  as long as the total occupation 
 number $n$ is large and coupling $\alpha$ is sufficiently weak. 
  The reason is that the species are only distinguished by the quantum number $j$ that is associated with the weak coupling.  \\

To reiterate, if $n$ is large, 
the state $\ket{f}$ is essentially
classical, even if the individual occupation numbers are minimal, 
$n_j =1$. 
This is because an observer (Alice) needs a very long time 
in order to distinguish the individual ``colors" of the constituents if 
their quantum coupling $\alpha$ is extremely weak, 
$\alpha = {1 \over N}$. Indeed, imagine that Alice wishes to distinguish
the state $\ket{f}$ with $n_1=N, n_{j\neq 1} =0$ from the one with 
$n_1 =n_2=...=n_N=1$. 
In order to read-out the color content of the state $\ket{f}$, Alice has 
to initiate  an act of interaction between the individual 
$a$-quanta and some color-sensitive external probe.
 However, the rate of such interaction is suppressed by powers of $\alpha$. Correspondingly, the minimal time-scale required for the
 measurement per particle is $t \propto \frac{1}{\alpha}$.    
 Thus, the detection of the species quantum identities demands an investment of a macroscopically-long time-scale.  
 On the shorter time-scales, the only observable effects are the collective 
 $N$-particle processes that are controlled by the 't Hooft coupling
 $\lambda_t = \alpha N$. The latter effects do not vanish 
 in the 't Hooft's large-$N$ limit, and therefore,  
 are {\it classically-observable}.

\end{document}